\documentclass[11pt]{article}
\usepackage{amssymb}
\usepackage{amsthm}
\usepackage{amsmath}
\usepackage{graphicx}
\usepackage{fullpage}
\usepackage{subcaption}
\usepackage{graphics}
\usepackage{tikz}
\usepackage{color}
\usepackage{listings}
\usepackage{fancybox}
\usepackage{complexity}
\usepackage{microtype}
\usepackage[inline]{enumitem}

\newtheorem{theorem}{Theorem}[section]
\newtheorem{lemma}[theorem]{Lemma}

\newtheorem{corollary}[theorem]{Corollary}
\newtheorem{definition}[theorem]{Definition}

\newtheorem{remark}[theorem]{Remark}

\newtheorem{folklore}[theorem]{Folklore}

\DeclareMathOperator*{\argmax}{arg\,max}

\DeclareMathOperator{\maj}{Maj}

\newcommand*{\triclusp}[3]{temporal $({#1}$,${#2}$,${#3})$-clustering}
\newcommand*{\TRICLUSp}[3]{\textsc {Temporal $({#1}$,${#2}$,${#3})$-Clustering}}
\newcommand*{\triappp}[3]{$({#1}$,${#2}$,${#3})$-approximation}

\def\CNF/{\textsc{CNF}}

\def\DKC/{{\sc Temporal $k$-Center}}
\def\DRDS/{{\sc Temporal $r$-Dominating Set}}
\def\KC/{{\sc $k$-Center}}
\def\RDS/{{\sc $r$-Dominating Set}}
\def\SETC/{{\sc Set-Cover}}
\def\NIL/{{\sc Nil}}

\def\TKMED/{\textsc {Temporal $k$-Median}}
\def\TKMEDC/{\textsc {Temporal $(k$,$r$,$\delta)$-Median Clustering}}
\def\TKMEANS/{\textsc {Temporal $k$-Means}}
\def\TKMEANSC/{\textsc {Temporal $(k$,$r$,$\delta)$-Means Clustering}}
\def\TKC/{\textsc {Temporal $k$-Center}}
\def\TRUE/{\textsc {True}}
\def\FALSE/{\textsc {False}}
\def\YES/{\textsc {Yes}}
\def\NO/{\textsc {No}}

\def\triclus/{\triclusp{k}{r}{\delta}}
\def\TRICLUS/{\TRICLUSp{k}{r}{\delta}}
\def\GAP/{\GAPp{k'}{r'}{\delta'}}
\def\triapp/{\triappp{k'}{r'}{\delta'}}

\def\TC/{\textsc {Temporal Clustering}}
\def\MAX2SAT/{\textsc {MAX-$2$-$\SAT$}}
\def\THREESAT/{\textsc {$3$-$\SAT$}}

\def\TRUE/{\textsc {True}}
\def\FALSE/{\textsc {False}}

\newcommand*{\opt}[1]{{#1}_{\textsc{OPT}}}
\newcommand{\eps}{\varepsilon}

\DeclareMathOperator*{\diam}{diam}
\newcommand{\cost}{\mathsf{cost}}
\newcommand{\tube}{\mathsf{tube}}
\newcommand{\ball}{\mathsf{ball}}

\newcommand{\uncovered}[2]{\mathsf{uncovered(#1,#2)}}
\newcommand{\rad}{\ensuremath{\mathsf{rad}_{\infty}}}
\newcommand{\mediancost}{\ensuremath{\mathsf{rad}_1}}
\newcommand{\meancost}{\ensuremath{\mathsf{rad}_2}}
\newcommand{\nil}{\ensuremath{\mathsf{nil}}}
\newcommand{\head}{\ensuremath{\mathsf{head}}}
\newcommand{\tail}{\ensuremath{\mathsf{tail}}}
\newcommand{\val}{\ensuremath{\mathsf{val}}}

\theoremstyle{plain}
\title{Temporal Clustering} %\footnote{This work was partially supported by someone.}}
\author{
  Tamal K. Dey
  \thanks{
    Dept.~of Computer Science and Engineering, and Dept.~of Mathematics,
    The Ohio State University. Columbus, OH, 43201.
    \texttt{tamaldey@cse.ohio-state.edu}
  }
\and
  Alfred Rossi
  \thanks{
    Dept.~of Computer Science and Engineering,
    The Ohio State University. Columbus, OH, 43201.
    \texttt{rossi.49@osu.edu}
  }
\and
  Anastasios Sidiropoulos
  \thanks{
    Dept.~of Computer Science and Engineering, and Dept.~of Mathematics,
    The Ohio State University. Columbus, OH, 43201.
    \texttt{sidiropoulos.1@osu.edu}
  }
}

\begin{document}
\maketitle
\begin{abstract}
We study the problem of clustering sequences of unlabeled point sets taken from
a common metric space. Such scenarios arise naturally in applications where a
system or process is observed in distinct time intervals, such as biological
surveys and contagious disease surveillance. In this more general setting
existing algorithms for classical (i.e.~static) clustering problems are not
applicable anymore.

We propose a set of optimization problems which we collectively refer to as
\emph{temporal clustering}. The quality of a solution to a temporal clustering
instance can be quantified using three parameters: the number of clusters $k$,
the spatial clustering cost $r$, and the maximum cluster displacement $\delta$
between consecutive time steps. We consider spatial clustering costs which
generalize the well-studied $k$-center, discrete $k$-median, and discrete $k$-means
objectives of classical clustering problems. We develop new algorithms that
achieve trade-offs between the three objectives $k$, $r$, and $\delta$. Our
upper bounds are complemented by inapproximability results.
\end{abstract}

\tableofcontents

\section{Introduction}
Clustering points in a metric space is a fundamental problem that can be used to express a
plethora of tasks in machine learning, statistics, and engineering, and has been
studied extensively both in theory and in practice
\cite{arthur2007k,cabello2011geometric,forgy1965cluster,DBLP:journals/dcg/Har-PeledK07,har2004coresets,har2005fast,hochbaum1985best,jain1988algorithms,kanungo2002local,kolliopoulos2007nearly,lloyd1982least,thorup2001quick}.
Typically, the input consists of a set $P$ of points in some metric space and
the goal is to compute a partition of $P$ minimizing a certain objective, such
as the number of clusters given a constraint on their diameters.

We study the problem of clustering {\it sequences} of {\it unlabeled} point sets
taken from a common metric space. Our goal is to cluster the points in each
`snapshot' so that the cluster assignments remain coherent across successive
snapshots (across time). We formulate the problem in terms of tracking the {\it
centers} of the clusters that may merge and split over time while satisfying
certain constraints. Such instances are common in the study of time-evolving
processes and phenomena under discrete observation. As an example consider a
hypothetical study which aims to track the spread of a certain genetic mutation
in plants. Here, data collection efforts center on annual field surveys in which
a technician collects and catalogs samples. The location and number of mutation
positive specimens change from year to year. Clustering such spaces is clearly a
generalization of classical (static) clustering, which we refer to as
\emph{temporal clustering}. In this dynamic variant of the problem, apart from
the number of clusters and their radii, we also wish to minimize the extent by
which each cluster moves between consecutive snapshots.

\paragraph*{Related work}
Clustering of moving point sets has been studied in the context of \emph{kinetic
clustering}
\cite{Agarwal:2002,Basch1997:KDS,Guibas1998:KDS,har2004clustering,Gao2006:KDS,Friedler2010:KDS,Abam2009:KDS}.
In that setting points have identities (labels) which are fixed throughout their
motion, the trajectories of the points are known beforehand, and the goal is to
design a data structure which can efficiently compute a near-optimal clustering
for any given time step. In our setting, since the points are not labeled there
is, {\it a priori}, no explicit motion. Instead we are given a sequence of unlabeled points in a
metric space and are required to assign the points of each to a limited
number of temporally coherent clusters. Motion emerges as a consequence of cluster
assignment. Consequently, kinetic clustering algorithms cannot be used in our
setting.
Another related problem concerns clustering time series
under the Fr\'{e}chet distance \cite{driemel2016clustering}, with the clusters
being constrained to move along polygonal trajectories of bounded complexity.
This constraint is used to avoid overfitting, and is conceptually similar to our
requirement that the clusters remain close between snapshots.

\subsection{Problem formulations}
Let us now formally define the algorithmic problems that we study in this paper.
Perhaps surprisingly, very little is known for temporal clustering problems.
There are of course different optimization problems that one could define; here
we propose what we believe are the most natural ones.

\begin{figure}%
\centering
\begin{subfigure}[b]{0.3\linewidth}
  \includegraphics[width=\linewidth]{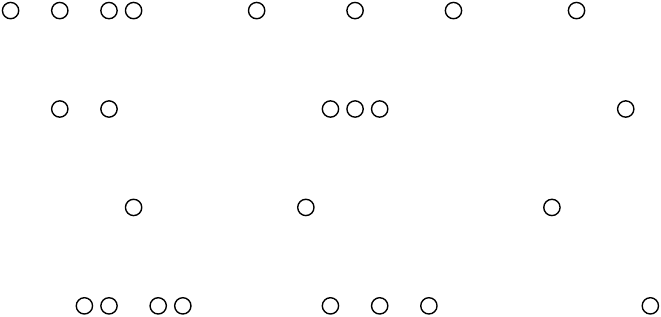}
  \caption{$P$\label{fig:diagram-p}}%
\end{subfigure}\hfill
\begin{subfigure}[b]{0.3\linewidth}
  \includegraphics[width=\linewidth]{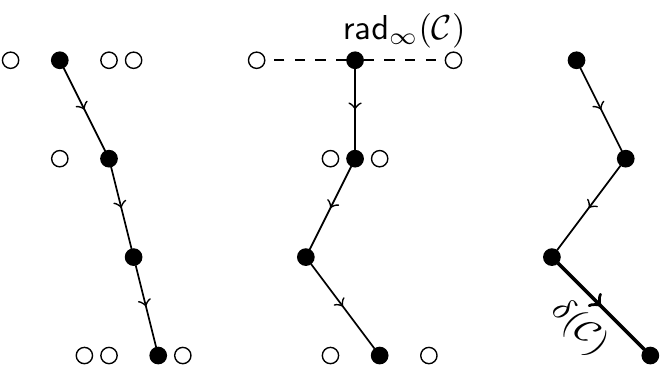}
  \caption{A temporal $3$-clustering\label{fig:diagram-p-k-3}}%
\end{subfigure}\hfill
\begin{subfigure}[b]{0.3\linewidth}
  \includegraphics[width=\linewidth]{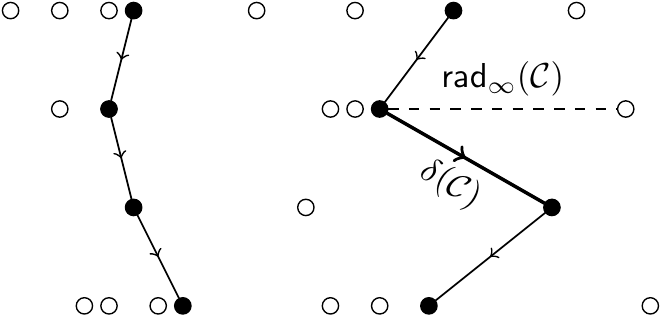}
  \caption{A temporal $2$-clustering\label{fig:diagram-p-k-2}}%
\end{subfigure}
\caption{%
    (\ref{fig:diagram-p}) A temporal-sampling $P$ of length $4$ where
        $P(i) \subset \mathbb{R}$ is drawn horizontally. Each level of $P$ is
        depicted as a row starting from $P(1)$ at the top.
    (\ref{fig:diagram-p-k-3}) The temporal-sampling $P$ shown with a clustering
        ${\cal C}$, consisting of $3$ clusters. The centers of each of $3$
        trajectories are depicted as filled circles in each level. Arrows are
        drawn between $\tau_j(i)$ and $\tau_j(i+1)$ for each trajectory
        $\tau_j$, $j \in \{1, 2, 3\}$. In the first level, a pair of points
        which achieve the spatial cost are joined to
        their respective cluster centers by a dashed edge. The arrow between the
        pair of centers which achieves maximum displacement is shown in bold.
    (\ref{fig:diagram-p-k-2}) The temporal-sampling $P$ shown with $2$ clusters.
      \label{fig:example-clusterings}
    }%
\end{figure}

We first define how the input to a temporal clustering problem is described. Let
$M=(X, d)$ be a metric space. Let $P(1), \ldots, P(t)$ be a sequence of $t$
finite, non-empty metric subspaces (points) of $M$. We refer to individual
elements of this sequence (the `snapshots') as \emph{levels}, and collectively
to $P$ as a \emph{temporal-sampling of $M$} of \emph{length} $t$. The
\emph{size} of $P$ is the total number of points over all levels, that is
$\sum_{i \in [t]} |P(i)|$. Let $\{\tau(i)\}_{i=1}^t$ be a sequence of points
such that $\tau(i) \in P(i)$ is a single point. We say that $\tau$ is a
\emph{trajectory} of $P$, and we let ${\cal T}(P)$ denote the set of all
possible trajectories of $P$. For some ${\cal C} \subseteq {\cal T}(P)$, we
denote by ${\cal C}(i)$ the set of points of the trajectories in ${\cal C}$
which lie in $P(i)$. In other words, ${\cal C}(i) = \bigcup_{\tau \in {\cal
C}}{\tau(i)}$. The set of trajectories ${\cal C}$ induces a clustering on each
level $P(i)$ by assigning each $p \in P(i)$ to the trajectory $\tau \in {\cal
C}$ that minimizes $d(p,\tau(i))$. We refer to the points of $\mathcal{C}(i)$ as
the \emph{centers} of level $i$. Intuitively, this formulation allows points in
different levels of $P$ which are assigned to the same trajectory to be part of
the same cluster; see Figure~\ref{fig:example-clusterings}. Further, observe that
trajectories may overlap allowing clusters to merge and split implicitly; see Figure~\ref{fig:diagram-grp-with-c}. We refer to ${\cal C}$ as a
\emph{temporal-clustering} of $P$.

%(though, strictly speaking, the temporal-clustering is induced by ${\cal C}$).

%We note that our use of the term \emph{trajectory} differs from the standard
%literature (\emph{c.f.} \cite{Buchin,Gudmundsson}).
%Alfred: I will find some references to add from these guys ^^
%Here, trajectories are merely
%sequences of points which serve as cluster representatives. In particular, the
%point of the trajectory in a level $i$ represents some subset of points of
%$P(i)$.

We now formalize the clustering objectives. Our approach is to treat temporal
clustering as a multi-objective optimization problem where we try to find
a collection of trajectories such that their induced clustering ensures
three conditions: (i) points in the same cluster remain near between successive levels
(\emph{locality}), (ii) the restriction of the clustering to any single level
fits the shape of the data (\emph{spatial constraint}), and (iii) we do
not return excessively many clusters (\emph{complexity}).
To measure how far some trajectory $\tau$ jumps, we define its
\emph{displacement}, denoted by $\delta(\tau)$, to be
$
\delta(\tau) = \max_{i\in [t-1]} d(\tau(i), \tau(i+1)).
$
We also define the displacement of ${\cal C}$ to be $\delta({\cal C}) =
\max_{\tau \in \mathcal{C}} \delta(\tau).$
Finally, we consider three different objectives for the spatial cost, which
correspond to generalization of the $k$-center, $k$-median, and
$k$-means respectively. The first one, corresponding to $k$-center, is the
maximum over all levels of the maximum cluster radius; formally
$
\rad({\cal C}) = \max_{i\in [t]} \max_{p \in P(i)} d(p,{\cal C}(i)),
$
where $d(p,{\cal C}(i)) = \min_{\tau\in {\cal C}} d(p, \tau(i))$.
The second and third spatial cost objectives, which corresponding to discrete $k$-median,
and discrete $k$-means (respectively),
are defined to be
$\mediancost({\cal C}) = \max_{i\in [t]} \sum_{p \in P(i)} d(p,{\cal C}(i))$,
and
$\meancost({\cal C}) = \max_{i\in [t]} \sum_{p \in P(i)} d(p,{\cal C}(i))^2$.

\begin{definition}
Let $r\in \mathbb{R}_{\geq 0}$, $\delta \in \mathbb{R}_{\geq 0}$. We say that a set of trajectories
${\cal C}\subseteq {\cal T}(P)$ is a \emph{\triclus/} of $P$ if $\rad(\mathcal{C}) \leq r$,
$\delta({\cal C}) \leq \delta$, and $|\mathcal{C}| \leq k$. (See
Figure~\ref{fig:example-clusterings} for an example.)
We further define \emph{temporal $(k,r,\delta)$-median-clustering} and
\emph{$(k,r,\delta)$-means-clustering} analogously by replacing $\rad$ by
$\mediancost$ and $\meancost$ respectively.
\end{definition}

We now formally define the optimization problems that we study. In the case of
static clustering, a natural objective is to minimize the maximum cluster
radius, subject to the constraint that only $k$ clusters are used; this is the
classical {\sc $k$-Center} problem \cite{hochbaum1985best}. Another natural
objective in the static case is to minimize the number of clusters subject to
the constraint that the radius of each cluster is at most $r$, for some given
$r>0$; this is the {\sc $r$-Dominating Set} problem
\cite{haynes1998fundamentals}. Our definition of temporal clustering includes
the temporal analogues of {\sc $k$-Center} and {\sc $r$-Dominating Set} as
special cases.

\begin{definition}[\TRICLUS/ problem]\label{def:TRICLUS}
An instance of the \TRICLUS/ problem is a tuple $(M, P, k, r, \delta)$, where
$M$ is a metric space, $P$ is a temporal-sampling of $M$, $k\in \mathbb{N}$,
$r\in \mathbb{R}_{\geq 0}$, and $\delta \in \mathbb{R}_{\geq 0}$. The goal is to
decide whether $P$ admits a \triclus/.
\end{definition}

\begin{definition}[\TRICLUS/ approximation]\label{def:APPROXTRICLUS}
Given an instance of the \TRICLUS/ problem consisting of a tuple $(M, P, k, r,
\delta)$, a \emph{\triappp{\alpha}{\beta}{\gamma}} is an algorithm which either
returns a \triclusp{\alpha k}{\beta r}{\gamma \delta} of $P$, or correctly
decides that no \triclus/ exists. In general $\alpha$, $\beta$, and $\gamma$ can
be functions of the input.
\end{definition}
We analogously define the \TKMEDC/ problem and approximation, and the \TKMEANSC/
problem and approximation by replacing in Definitions \ref{def:TRICLUS} and
\ref{def:APPROXTRICLUS} $(\cdot, \cdot, \cdot)$-clustering by $(\cdot, \cdot,
\cdot)$-median-clustering and $(\cdot, \cdot, \cdot)$-means-clustering
respectively.

\subsection{Our contribution}
To the best of our knowledge, this is the first study of the above models of
temporal clustering. Our main contributions consist of polynomial-time
approximation algorithms for several temporal clustering variants, and hardness
of approximation results for others.

\textbf{Temporal clustering.}
We begin by discussing our results on \TRICLUS/. We first consider the problem
of minimizing $r$ and $\delta$ while keeping $k$ fixed. This is a generalization
of the static {\sc $k$-Center} problem. The classical greedy algorithm that
yields the $2$-approximation to the static case, which is known to be optimal
assuming $\text{P}\neq \text{NP}$ \cite{hochbaum1985best}, does not appear to be
applicable in the temporal setting. The reason is that a static solution cannot
account for the requirement that the cluster centers of the same trajectory
cannot be too far apart in consecutive levels. We present a polynomial-time
$(1,2,1+2\eps)$-approximation algorithm where $\eps = r / \delta$ using a
different method. More specifically, our result is obtained via a reduction to a
network flow problem. We show that the problem is \NP-hard to approximate to
within polynomial factors even if we increase the radius by a polynomial factor.
Formally, we show that it is \NP-hard to obtain a
$(1,\poly(n),\poly(n))$-approximation.

Next we consider the problem of minimizing the number of clusters $k$, while
fixing $r$ and $\delta$. This is a generalization of the static {\sc
$r$-Dominating Set} problem. We obtain a polynomial-time $(\ln
n,1,1)$-approximation algorithm. For the static case, the polynomial-time $\ln
n$-approximation algorithm follows by a reduction to the {\sc Set-Cover}
problem, and is known to be best-possible
\cite{Dinur14,lund1994hardness,feige1998threshold}. However, in the temporal
case, this reduction produces an instance of {\sc Set-Cover} of exponential
size. Thus, it does not directly imply a polynomial-time algorithm for \DRDS/.
We bypass this obstacle by showing how to run the greedy algorithm for {\sc
Set-Cover} on this exponentially large instance in polynomial-time, without
explicitly computing the {\sc Set-Cover} instance. We also argue that $(\ln n,
1, 1)$-approximation is best possible by observing that $((1-\eps)\ln n,
2-\eps',\cdot)$-approximation is \NP-hard for any $\eps,\eps'>0$.

We further present a result that can be thought of as a trade-off between the
above two settings by allowing both the number of clusters and the radius to
increase. More precisely, we obtain a polynomial-time
$(2,2,1+\eps)$-approximation algorithm where $\eps = r / \delta$.
Interestingly, we can show that obtaining a $(1.005,2-\eps,\poly(n))$-approximation is
\NP-hard.

The following summarizes the above approximation algorithms.
\begin{theorem}
\TRICLUS/ admits the following algorithms:
\begin{enumerate}[ref={\thetheorem.\arabic*},topsep=0pt,itemsep=0pt]
\item \triappp{1}{2}{1+2\eps} where $\eps = r / \delta$, \label{thm:DKC}
\item \triappp{\ln(n)}{1}{1}, \label{thm:DRDS-approx}

\item \triappp{2}{2}{1+\eps} where $\eps = r / \delta$, \label{thm:BICRIT-approx}
\end{enumerate}
where $n$ is the size of the temporal-sampling.
Moreover, the running time of all of these algorithms is $O(n^3)$.
\end{theorem}
We prove Theorems~\ref{thm:DKC},~\ref{thm:DRDS-approx},~\ref{thm:BICRIT-approx}
in Sections~\ref{sec:k-center},~\ref{sec:DRDS},~\ref{sec:BICRIT-approx},
respectively.

It is important that the approximation in displacements for Theorem~\ref{thm:DKC} and Theorem~\ref{thm:BICRIT-approx} takes into account the factor $\eps=r / \delta$ if a polynomial time algorithm is aimed for. This is because
our inapproximability results as summarized below show that the problem is NP-hard otherwise.
\begin{theorem}
The status of \TRICLUS/ with temporal-samplings of size $n$ is as follows:
\begin{enumerate}[ref={\thetheorem.\arabic*},topsep=0pt,itemsep=0pt]
\item There exist universal constants $c>0$, $c'>0$ such that
\triappp{1}{c n^{s(1-\eps)}}{c' n^{(1-s)(1-\eps)}} is \NP-hard for any
$\eps, s \in \mathbb{R}$ where $\eps > 0$ and $s \in [0,1]$.
\label{thm:1polypolyhard}
\item \triappp{(1-\eps)\ln(n)}{2-\eps'}{\cdot} is \NP-hard for any fixed $\eps>0$, $\eps' > 0$.
\label{thm:hardassetcover}
\item There exists a universal constant $c$ such that
\triappp{1.00579}{2-\eps'}{c n^{1-\eps}} is \NP-hard for any fixed $\eps>0$,
$\eps' > 0$. \label{thm:ccpolyhard}
\end{enumerate}
Moreover, items \ref{thm:1polypolyhard} and \ref{thm:ccpolyhard} remain \NP-hard
even for temporal-samplings in $2$-dimensional Euclidean space.
\end{theorem}
We discuss Theorem~\ref{thm:1polypolyhard} in
section~\ref{sec:nphardsmalldelta}. The discussion of
Theorem~\ref{thm:hardassetcover}, and Theorem~\ref{thm:ccpolyhard} are deferred
to section~\ref{sec:setchard}, and section~\ref{sec:inapprox2}, respectively.

\textbf{Temporal median clustering.}
We next discuss our result on the \TKMEDC/ problem. The static {\sc $k$-Median}
problem admits a $O(1)$-approximation via local search
\cite{arya2004local,korupolu2000analysis}. In Section~\ref{sec:tkmapprox} we
show that the local search approach fails in the temporal case, even on temporal
samplings of length two. We present an algorithm that achieves a trade-off
between the number of clusters and the spatial cost. The result is obtained via
a greedy algorithm, which is similar to the one used for the {\sc $k$-Set Cover}
problem. The result is summarized in the following theorem.

\begin{theorem}\label{thm:kmlogeps1}
%
%Let $(M, P, k, r, \delta)$ be an instance of the \TKMEDC/ problem.
%Let $n$ denote the size of $P$ and let $\Delta$ denote the spread of $M$.
For any fixed $\eps>0$, there exists a $(O(\log(n\Delta/\eps))$, $1+\eps,1)$-median-approximation algorithm with running time $\poly(n, \log(\Delta/\eps))$,
on an instance of size $n$ and a metric space of spread $\Delta$.
\end{theorem}

The result is obtained by iteratively selecting a trajectory
which minimizes a certain potential function.
The proof uses submodularity and monotonicity of the potential function.
These properties remain true if the potential function is modified by replacing $d(p, {\cal C}(i))$
with $d(p, {\cal C}(i))^2$, and thus an identical theorem holds for
\TKMEANS/.

We complement the above algorithm by showing the following hardness result.

\begin{theorem}
The status of \TKMEDC/ with temporal-samplings of size $n$ is as follows:
\begin{enumerate}[ref={\thetheorem.\arabic*},topsep=0pt,itemsep=0pt]
\item
There exist universal constants $c_r$, $c_\delta$ such that
\triappp{1}{c_rn^{s(1-\eps)}}{c_\delta n^{(1-s)(1-\eps)}} for \TKMED/ is \NP-hard
for any $\eps, s \in \mathbb{R}$ where $\eps > 0$ and $s \in [0,1]$.
\label{thm:1polypolyhardkm}

\item Let $c$, $s$ be the constants from Theorem~\ref{thm:max2sgap}. Let $0 \leq
f < c - s$. Then $(\frac{3-(s+f)}{3-c}, 1+c_rf, c_\delta
n^{1-\eps})$-approximation is \NP-hard for any fixed $\eps > 0$ and some
constants $c_r$, $c_\delta$. \label{thm:ccpolyhardkm}
\end{enumerate}
Moreover, item \ref{thm:1polypolyhardkm} remains hard even for
temporal-samplings from $2$-dimensional Euclidean space.
\end{theorem}
The clustering instances used in the proofs of Theorem~\ref{thm:1polypolyhardkm}
in section~\ref{sec:1polypolyhardkm} and Theorem~\ref{thm:ccpolyhardkm} in
section~\ref{sec:ccpolyhardkm} involve clusterings which use only a constant
number of points per cluster, thus the same constructions suffice to prove
hardness of \TKMEANSC/ with only slight modification of the distances.

\begin{figure}
\centering
\includegraphics[width=0.4\linewidth]{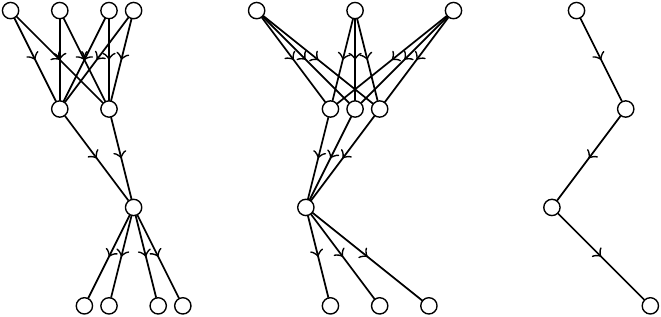}
\caption{
The graph $G_{\delta}(P)$ for $P$ from the previous diagram and some $\delta$.
Points which are within a distance of $\delta$ in adjacent levels are connected
by a directed edge which points toward the higher indexed level.\label{fig:succ}
}
\end{figure}

\textbf{Additional notation and preliminaries.}
Let $r>0$. An $r$-net in some metric space $(X,d)$ is some maximal $Y\subseteq
X$, such that for any $x,y\in Y$, with $x\neq y$, we have $d(x,y)>r$.
Let $P$ be a temporal-sampling of length $t$ in some metric space $(X,d)$. Let
$V(P,i)= \bigcup_{x\in P(i)} \{(i,x)\}$ for all $i \in [t]$. For any trajectory
$\tau$, and for any $r \geq 0$, the \emph{tube} around $\tau$ of \emph{radius}
$r$, denoted by $\tube(\tau,r)$, is defined to be
$\tube(\tau,r) = \bigcup_{i\in [t]} \{(i, x) \in V(P,i) \mid x \in \ball(\tau(i), r) \}$,
where for $x \in X$, $r \in \mathbb{R}_\geq{0}$, we use the notation
$\ball(x,r)$ to denote a closed ball of radius $r$. Let $\delta \in
\mathbb{R}_{\geq 0}$. The directed graph $G_{\delta}(P)$ has as vertices
$V(P,i)$ for all $i \in [t]$. For any $i \in [t-1]$ there is an edge between $p
\in V(P,i)$ and $q \in V(P,i+1)$ whenever $d(p,q) \leq \delta$ (see
Figure~\ref{fig:succ}).

\section{Algorithms}

\subsection{Exact number of clusters: \triappp{1}{2}{1+2\eps}}
\label{sec:k-center}
In this section, we consider the problem of computing a temporal clustering by
relaxing the radius and the displacement, while keeping the number of clusters
exact. This is a temporal analogue of the {\sc $k$-Center} problem. We first
present a polynomial time \triappp{1}{2}{1+2\eps} where $\eps = r / \delta$. In
section~\ref{sec:nphardsmalldelta} complement this with an inapproximability
result.

\begin{figure}%
\begin{subfigure}[b]{.48\textwidth}
\centering
  \includegraphics[width=0.75\linewidth]{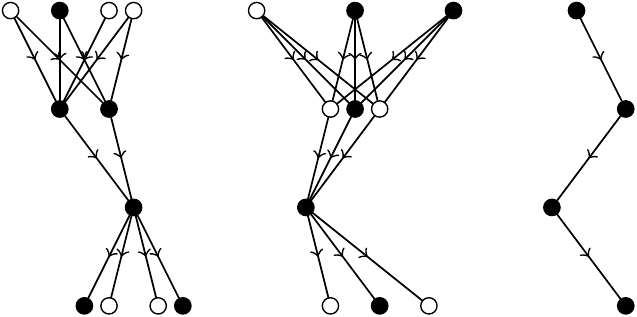}
  \vspace{4.2ex}
  \caption{The graph $G_\gamma(P)$.}%
  \label{fig:diagram-grp-with-c}
\end{subfigure}
\begin{subfigure}[b]{.48\textwidth}
  \centering
  \includegraphics[width=0.75\linewidth]{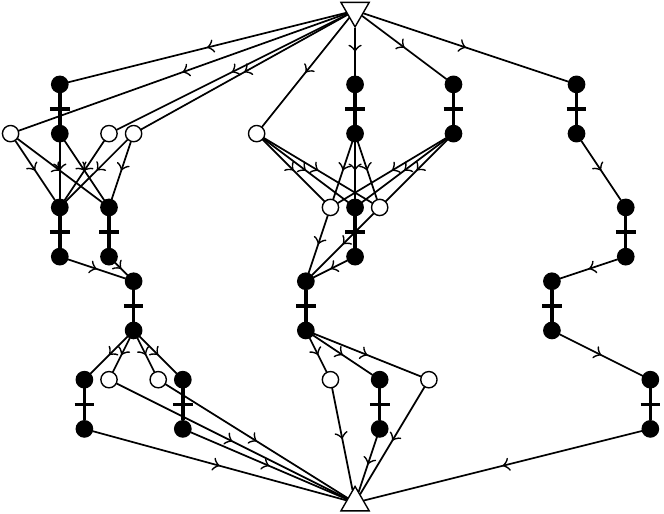}
  \caption{The flow network $N_\gamma(P,C)$.}%
  \label{fig:diagram-grp-with-c-flow}
\end{subfigure}
\caption{%
(\ref{fig:diagram-grp-with-c}) The graph $G_\gamma(P)$ for $P(i) \subset \mathbb{R}$ and some $\gamma>0$. The vertices in $C(i)$, $i\in [4]$, are indicated with filled circles.
(\ref{fig:diagram-grp-with-c-flow}) The flow network  $N_\gamma(P,C)$ corresponding to $G_{\gamma}(P)$. Every node from $C$ has been split into an edge.
}%
\end{figure}

\textbf{An auxiliary network flow problem.}
The high-level idea of the polynomial time algorithm is to use a reduction  to a
specific network flow problem. Specifically, we seek a minimum flow which
satisfies lower bound constraints along certain edges. This is the so-called
minimum flow, or minimum feasible flow problem
\cite{Aho87efficientalgorithms,GT89}. We now
formally define this flow network. For each $i \in [t]$, let $C(i) \subseteq
P(i)$. Let $\rho>0$. We construct a flow network, denoted by $N_{\gamma}(P,C)$
where $C$ is the sequence of centers $C(i)$ for $i \in [t]$. We start with the
graph $G_{\gamma}(P)$. In level $i$, we replace each vertex $v=(i,c)$ for $c \in
C(i)$ by a pair of vertices $\tail(v)$ and $\head(v)$, and we connect them by an
edge $(\tail(v), \head(v))$. For vertices $v=(i,p)$ where $p \in P(i) \setminus
C(i)$ we define $\tail(v)=\head(v)=v$. Now for \emph{any} vertex $v$, all
incoming edges to $v$ become incoming edges to $\tail(v)$, and all outgoing
edges from $v$ become outgoing edges from $\head(v)$. We add a source vertex $s$
and a sink vertex $s'$. For all $p\in P(1)$, we add an edge from $s$ to
$\tail((1,p))$. Similarly, for all $p\in P(t)$, we add an edges from
$\head((t,p))$ to $s'$. We set the capacity of each edge to be $\infty$.
Finally, we set a lower bound of $1$ to the capacity of every edge
$(\tail(v),\head(v))$, for all $v=(i,c)$, $c\in C(i)$, $i\in [t]$ (see Figure
\ref{fig:diagram-grp-with-c-flow}).

%We are now ready to describe the algorithm.
\textbf{Algorithm.} We first compute a net at every
level of the temporal-sampling and then we reduce the problem of computing a
temporal clustering to a flow instance, using the network flow defined above. By
computing an integral flow and decomposing it into paths, we obtain a collection
of trajectories. The lower bound constraints ensure that all net points are
covered; this allows us to show that all points are covered by the tubular
neighborhoods of the trajectories.
Formally, the algorithm consists of the following steps:
\begin{description}[topsep=0pt,itemsep=0pt]
\item{\textbf{Step 1: Computing nets.}} For each $i\in [t]$, compute a $2r$-net
$C(i)$ of $P(i)$. If for some $i\in [t]$, $|C(i)| > k$, then return \nil.
\item{\textbf{Step 2: Constructing a flow instance.}} We construct the minimum
flow instance $N_{2r+\delta}(P,C)$.
\item{\textbf{Step 3: Computing a collection of trajectories.}} If the flow
instance $N_{2r+\delta}(P,C)$ is not feasible, then return $\nil$. Otherwise,
find a minimum integral flow $F$ in $N_{2r+\delta}(P,C)$, satisfying all the
lower bound constraints. Decompose $F$ into a collection of paths, each carrying
a unit of flow. The restriction of each path in $G$ is a trajectory. Output the
set of all these trajectories.
\end{description}
%This concludes the description of the algorithm.

%\paragraph*{Analysis}
%
Throughout the rest of this section let $P$ be a temporal-sampling. We now show
that if there exists a \triclus/, then the above algorithm outputs a temporal
$(k,2r,(1+2\eps)\delta)$-clustering where $\eps = r / \delta$.

\begin{lemma}\label{lem:injective}
Suppose that $P$ admits a \triclus/, $\mathcal{Q}$. For each $i \in [t]$ let
$Q(i)$ denote the level $i$ centers of $\mathcal{Q}$, and let $C(i)$ be a
$2r$-net of $P(i)$. Then the map $\pi_i: C(i)\rightarrow Q(i)$ which sends each
$2r$-net center to a nearest center in $Q(i)$ is injective.
\end{lemma}

\begin{proof}
First, observe that for each $c \in C(i)$, $d(c,\pi_i(c)) \leq r$ because
$r$-balls centered at the points in $Q(i)$ cover $P(i)$ and hence $C(i)$. For
injectivity of $\pi_i$, observe that, $\pi_i(c)\not=\pi_i(c')$ for $c\neq c'$
because otherwise the inequality $
d(c,c')\leq d(c,\pi_i(c))+d(c',\pi_i(c')) \leq 2r
$ holds violating the property that $C(i)$ is a $2r$-net.
\end{proof}

Since for each $i \in [t]$, the map $\pi_i$ is injective, it follows that
$|C(i)| \leq |Q(i)| \leq k$. So, we have the following immediate Corollary.

\begin{corollary}\label{lem:netsize}
If $P$ admits a \triclus/ then for any $i \in [t]$, any $2r$-net $C(i)$ of
$P(i)$ has $|C(i)|\leq k$.
\end{corollary}

\begin{lemma}\label{lem:flow}
If $P$ admits a \triclus/ then for any level-wise $2r$-net $C$, the flow
instance $N_{2r+\delta}(P,C)$ admits a feasible flow of value $k$.
\end{lemma}

\begin{figure}
\centering{\includegraphics[width=0.6\linewidth]{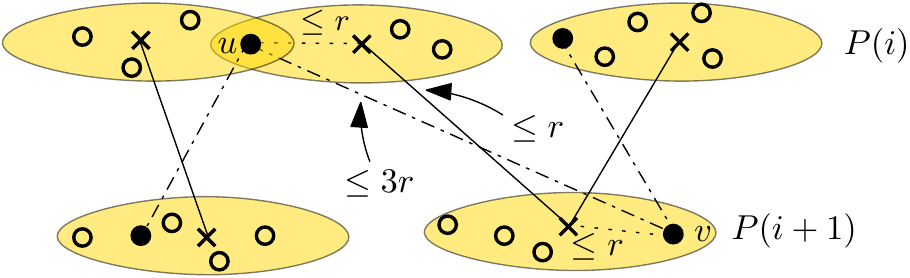}}
\caption{The crosses, filled circles, and empty circles are optimal centers, net
points, and other points respectively. The paths ($\tau$) in optimal solution
and the deformed paths($\tau'$) are indicated with solid and dotted edges
respectively.\label{fig:3approx}}
\end{figure}

\begin{proof}
Fix a \triclus/ $\mathcal{Q}$ and let $\tau$ denote one of its $k$ trajectories.
The graph $G_{2r+\delta}(P)$ contains a path corresponding to $\tau$ as the
distance between any pair of consecutive points in $P$ is at most $\delta$. For
each $i$, let $\pi_i:C(i)\rightarrow Q(i)$ denote a map which sends each
$2r$-center of $C(i)$ to a nearest center in $Q(i)$. We modify $\tau$ to produce
some path $\tau'$ in $G_{2r+\delta}(P)$ as follows: for every level $i$ such
that $\tau(i)=\pi_i(c_i)$ for some net-point $c_i \in C(i)$ we let $\tau'(i) =
c_i$, otherwise we set $\tau'(i) = \tau(i)$. We observe that in the worst case
the distance between consecutive points, say $u=\tau'(i)$ and $v=\tau'(i+1)$, is
at most $2r+\delta$ because of the following inequality (see
Figure~\ref{fig:3approx})
$
d(u,v)\leq d(u,\tau(i))+d(\tau(i),\tau(i+1)) + d(\tau(i+1), v) \leq r + \delta + r
$.
It follows that $\tau'$ is indeed a path in $G_{2r+\delta}(P)$.
Further, by the injectivity of each map $\pi_i$ (Lemma~\ref{lem:injective})
which is used in deforming $\tau$ to $\tau'$, we have that for every net point,
there exists some $\tau'$ that contains it. In other words, all net points
$C(i)$ are covered by the paths $\tau'$. For each optimal trajectory $\tau$, let
$\tau''$ be the path in $N_{2r+\delta}(P,C)$ obtained from $\tau'$ by connecting
$s$ to the first vertex in $\tau'$, and the last vertex in $\tau'$ to $t$. By
routing a unit of flow in $N_{2r+\delta}(P,C)$ along each such $\tau''$ we
obtain a flow of value at most $k$ that meets all the demands along the edges
corresponding to net points $C$, concluding the proof.
\end{proof}

\begin{lemma}\label{lem:DKC}
Given $k$, $r$, $\delta$, and a temporal-sampling $P$, with $|P|=n$, there
exists an $O(n^3)$-time algorithm that either correctly decides $P$ does not
admit a \triclus/, or outputs some \triclusp{k}{2r}{2r+\delta}.
\end{lemma}
\begin{proof}
Lemmas \ref{lem:netsize} and \ref{lem:flow} imply that if a \triclus/ exists,
then the algorithm does not return \nil, and thus outputs a set $T$ of at most
$k$ trajectories. Let ${\cal C}$ be the temporal clustering corresponding to
$T$. Each trajectory in $T$ corresponds to a path in $G_{2r+\delta}(P)$, thus
has displacement at most $2r+\delta$. Therefore $\delta({\cal C}) \leq
2r+\delta$. Since $F$ is a feasible flow, it follows that all lower bound
constraints in $N_{2r+\delta}(P,C)$ are satisfied. Thus for all $i\in [t]$, for
all $c\in C(i)$, there exists at least one unit of flow along the edge
$(\tail(v), \head(v))$ corresponding to the vertex $v=(i,c)$; it follows that
there exists some trajectory containing $c$ in level $i$. Since for all $i\in
[t]$, $C(i)$ is a $2r$-net of $P(i)$, it follows that $P(i) \subseteq
\bigcup_{c\in C(i)} \ball(c, 2r)$. Thus $\bigcup_{i\in [t]} V(P,i) \subseteq
\bigcup_{\tau \in T} \tube(\tau, 2r)$, which implies that $\rad({\cal C}) \leq
2r$. We thus obtain that ${\cal C}$ is a \triclusp{k}{2r}{2r + \delta}.
Finally, we bound the running time. Computing the $2r$-nets over all levels,
checking their sizes can be done in $O(nk)$ time. Building $G_{2r+\delta}(P)$
and $N_{2r+\delta}(P,C)$ can be done in $O(n^2)$ time. Finding an integral
solution to $N_{2r+\delta}(P,C)$ takes $O(n^3)$ time using the algorithm of
Gabow and Tarjan \cite{GT89}. Decomposing the resulting flow takes $O(n^3)$
time. We conclude that the entire procedure completes in $O(n^3)$ time.
\end{proof}

Writing $\eps = r / \delta$, we immediately obtain Theorem~\ref{thm:DKC}
from Lemma~\ref{lem:DKC}.

\subsection{Exact radius and displacement: \triappp{\ln(n)}{1}{1}}
\label{sec:DRDS}
In this section we consider the case where the number of clusters is allowed to
be approximated in analogy to the static $r$-Dominating Set problem. We present
a polynomial-time \triappp{\ln(n)}{1}{1} algorithm. In
Section~\ref{sec:setchard} we argue that this result is tight in the sense that
obtaining a \triappp{(1-\eps)\ln(n)}{1}{1} is \NP-hard for any fixed $\eps > 0$.

Let $P$ be a
temporal-sampling of length $t$. For any $\delta \geq 0$, we denote by
$\mathcal{T}_\delta(P)$ the set of all trajectories of displacement at most
$\delta$.
Given an instance of the \TRICLUS/ problem consisting of a tuple $(M,
P, k, r, \delta)$, the high level idea is to express the problem as an instance
of \SETC/. Recall that an instance of \SETC/ consists of a pair $(U, {\cal S})$,
where $U$ is a set, and ${\cal S}$ is a collection of subsets of $U$. The goal
is to find some ${\cal S}'\subseteq {\cal S}$, minimizing $|{\cal S}'|$, such
that $U\subseteq \bigcup_{X\in {\cal S}'} X$, if such ${\cal S}'$ exists. We set
$ U = \bigcup_{i\in [t]} V(P,i), $
and
$ {\cal S} = \bigcup_{\tau \in \mathcal{T}_\delta(P)} \{\tube(\tau, r)\}. $
We will show that a solution to the \SETC/ instance $(U,{\cal S})$ can be used
to obtain a \triclusp{\ln(n) k}{r}{\delta}. Note that ${\cal S}$ can
have cardinality exponential in the size of the input. However, as we shall see,
we can still obtain an approximate solution for $(U,{\cal S})$ in
polynomial-time.

We first establish that any $\alpha(n)$-approximate solution to $(U, {\cal S})$
can be converted, in polynomial-time, to a \triclusp{\alpha(n) k}{r}{\delta}.
Let $\opt{s}$ denote the minimum cardinality of any feasible solution for $(U,
{\cal S})$ when it exists. Similarly, let $\opt{k}$ denote the smallest value of
$k'$ such that $P$ admits a \triclusp{k'}{r}{\delta}.

\begin{lemma}\label{lem:opt_k_opt_s}
$\opt{k}=\opt{s}$.
\end{lemma}

\begin{proof}[Proof of Lemma~\ref{lem:opt_k_opt_s}]
Let ${\cal C}=\{c_i\}_{i=1}^t$ be a \triclusp{k'}{r}{\delta} of $P$, for some
$k'\in \mathbb{N}$. Let $\tau^1,\ldots,\tau^{k'}$ be the natural decomposition
of ${\cal C}$ into a collection of trajectories, where
$\tau^j=c_1(j),\ldots,c_t(j)$, for all $i\in [k']$. Let ${\cal S}'=\bigcup_{j\in
[k']} \{\tube(\tau^j, r)\}$. Since $\delta({\cal C})\leq \delta$, it follows
that $\delta(\tau^j)\leq \delta$, for all $j\in [k']$. Thus, $\tau^j\in {\cal
T}_\delta(P)$, for all $j\in [k']$, which implies that ${\cal S}'\subseteq {\cal
S}$. Since ${\cal C}$ is a \triclusp{k'}{r}{\delta}, it follows that for all
$i\in [t]$, for all $x\in P(i)$, there exists some $j\in [k']$ such that
$d(\tau^j(i),x)\leq r$. It follows that
\[ U = \bigcup_{i\in [t]} V(P,i) \subseteq \bigcup_{j\in [k']} \tube(\tau^j,r) =
\bigcup_{X\in {\cal S}'} X. \]
We have established that ${\cal S}'$ is a valid solution for $(U,{\cal S}')$,
with $|{\cal S}'|=k'$, which implies that $\opt{s}\leq \opt{k}$.

Conversely, let ${\cal S}'\subseteq {\cal S}$ be a solution for $(U, {\cal S})$.
Let $I = \{\tau \in {\cal T}_\delta(P) : \{\tube(\tau,r)\}\in {\cal S}'\}$. Fix
an arbitrary ordering $I=\{\tau^1,\ldots,\tau^{|{\cal S}'|}\}$. We may now
define the clustering ${\cal C}=\{c_i\}_{i\in [t]}$, where $c_i(j)=\tau^j(i)$,
for all $i\in [t]$, $j\in [|{\cal S}'|]$. Since ${\cal S}'$ is a feasible
solution for $(U,{\cal S})$, it is immediate that ${\cal S}$ is a
\triclusp{|{\cal S}'|}{r}{\delta}, which implies that $\opt{k}\leq \opt{s}$.
\end{proof}
We next establish the following result which
allows us to run the greedy algorithm for \SETC/ on the instance $(U,{\cal S})$
in polynomial-time, even though $|{\cal S}|$ can be exponentially large.

\begin{lemma}\label{lem:find_path}
Let ${\cal S}'\subsetneq {\cal S}$. There exists an $O(n^2)$ time algorithm
which computes some $X\in {\cal S}\setminus {\cal S}'$, maximizing $\left|X\cap
\left(U\setminus \bigcup_{Y\in {\cal S}'} Y\right)\right|$. Moreover, the
algorithm outputs some trajectory $\tau\in {\cal T}_{\delta}(P)$, such that
$X=\{\tube(\tau,r)\}$.
\end{lemma}

\begin{proof}[Proof of Lemma~\ref{lem:find_path}]
We first establish some notation. Let $G=G_\delta(P)$. Note that due to the
orientation of the edges, any path $Q$ in $G$ has at most one vertex at level
$i$ for any $i \in [t]$. Recall that this vertex is of the form $(i, x)$ for
some $x \in P(i)$. For convenience we let $Q(i)=x$ whenever such a vertex
exists. Otherwise, we say that $Q(i)=\nil$. We extend the notion of a tube to
paths in $G$, by defining
\[ \tube(Q,r) = \bigcup_{i\in [t] : Q(i)\neq \nil} \{(i, x) \in V(P,i) \mid x \in \ball(Q(i), r) \} \]
We also denote those elements of $U$ which are not covered by ${\cal S}'$ as
$\uncovered{U}{{\cal S}'}$. In other words, $ \uncovered{U}{{\cal
S}'}=U\setminus \bigcup_{Y\in {\cal S}'} Y. $

The algorithm computes the desired $X$ via dynamic programming, as follows. The
dynamic programming table is indexed by $U$. For each $(i, x) \in U$, we compute
a path $Q$ in $G$ that we store at location $(i,x)$ of the table. More
precisely, if there is no path from $(i, x)$ to some vertex in $(t, y)$, where
$y \in P(t)$, then we set $Q=\nil$. Otherwise, we set $Q$ to be a path that
starts from $(i, x)$, and terminates at some vertex $(t, y)$, with $y \in P(t)$,
maximizing the quantity
$ \val(Q) = \left|\tube(Q,r) \cap \uncovered{U}{{\cal S}'} \right|. $
For every $x\in P(t)$, the only choice for $Q$ is $Q=x$. Thus we may fill in the
entries of the table indexed by $(t,x)$, for all $x\in P(t)$. Next, for each
$i\in [t-1]$, for each $x\in P(i)$, we compute the path $Q$ to be stored at
location $(i,x)$, assuming that all the entries indexed by $(i+1,y)$, for all
$y\in P(i+1)$ have already been computed. It is immediate for each $Q$ that
starts from $x$ and terminates at $P(t)$, we have
$ \val(Q) = \left|\ball(x,r) \cap \uncovered{U}{{\cal S}'} \right| + \val(Q'), $
where $Q'$ is the suffix of $Q$ obtained after removing $x$. Thus, in order to
compute the desired path $Q$ for $x$ that maximizes $\val(Q)$, it suffices to
find the path $Q'$ that maximizes $\val(Q')$, and starts from some neighbor of
$x$ in $P(i+1)$. This completes the description of the algorithm for filling in
the values of the dynamic table. We may now set $X=\tube(Q^*, r)$, where $Q^*$
is the path stored in the entry $(1,x)$, for some $x$ that maximizes
$\val(Q^*)$.
We now show how to complete this procedure in $O(n^2)$ time. During a
precomputation phase we can construct $\uncovered{U}{{\cal S}'}$ in $O(n)$ time.
Further, we can check whether an uncovered point intersects the $\ball(x, r)$ in
constant-time. This allows us to precompute and store $\left|\ball(x,r) \cap
\uncovered{U}{{\cal S}'} \right|$ for each node $(i, x) \in V(G)$ in time linear
time, for a total of $O(n^2)$ over all nodes. Now note that in populating the
table, the algorithm visits each node in $V(G)$ and evaluates the choice of
taking the path associated with each of its successors in $G$ as a suffix. By
also keeping the value of a path in the table this decision can be made in
constant time using only stored or precomputed information. Thus the algorithm
takes $O(n^2)$ time over all.
\end{proof}

We are now ready to prove
Theorem~\ref{thm:DRDS-approx}.

\begin{proof}[Proof of Theorem~\ref{thm:DRDS-approx}]
Recall that the classical greedy algorithm for \SETC/ computes a solution ${\cal
S}'\subseteq {\cal S}$, if one exists, as follows: Initially, we set ${\cal
S}'=\varnothing$. At every iteration, we pick some $X\in {\cal S}\setminus {\cal
S}'$ such that $\left|X\cap \left(U\setminus \bigcup_{Y\in {\cal S}'}
Y\right)\right|$ is maximized, and we add $X$ to ${\cal S}$. The algorithm stops
when either $U$ is covered by ${\cal S}$, or when no further progress can be
made, i.e.~when $\left|X\cap \left(U\setminus \bigcup_{Y\in {\cal S}'}
Y\right)\right|=0$; in the latter case, the instance $(U,{\cal S})$ is
infeasible. It is well-known that this algorithm achieves an approximation ratio
of $\ln n$ for \SETC/ \cite{johnson1974approximation}.
Now if $(U,{\cal S})$ is infeasible the above procedure detects this and
terminates. Otherwise, let ${\cal S}'\subseteq {\cal S}$ be the feasible
solution found by repeatedly using the procedure described in
Lemma~\ref{lem:find_path}. The corresponding trajectories returned by this
procedure form a \triclusp{k'}{r}{\delta} of $P$, for some $k'=|{\cal S}'| \leq
\ln n \cdot \opt{s}$. By Lemma~\ref{lem:opt_k_opt_s} it follows that $k' \leq
\ln n \cdot \opt{k} \leq \ln n \cdot k$. Thus we obtain an
\triappp{\ln(n)}{1}{1}.
Finally, to bound the running time note that in the worst case, the total number
of calls to the procedure in Lemma~\ref{lem:find_path} is $n$ since at every
step we cover at least uncovered point. The theorem now follows by the fact that
each call takes $O(n^2)$ time.
\end{proof}

\subsection{Approximating all parameters: \triappp{2}{2}{1+\eps}}
\label{sec:BICRIT-approx}
So far we have constrained either the number of clusters or the radius and the
displacement to be exact. We now describe an algorithm that relaxes all three
parameters simultaneously. We present a polynomial-time
$(2,2,1+\eps)$-approximation algorithm where $\eps = r / \delta$. We complement
this solution later in section~\ref{sec:inapprox2} by showing that it is
\NP-hard to obtain a $(1.005,2-\eps,\poly(n))$-approximation for any $\eps>0$.

\begin{lemma}\label{lem:flow_2r}
If $P$ admits a \triclus/ then for any level-wise $2r$-net $C$, the flow
instance $N_{r+\delta}(P,C)$ admits a feasible flow of value $2k$.
\end{lemma}

\begin{figure}
\centering
\begin{subfigure}[c]{.3\textwidth}
  \centering
  \includegraphics[width=\linewidth]{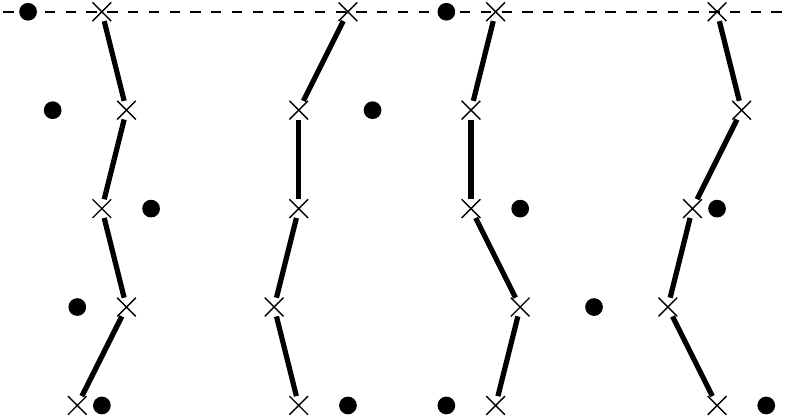}
  \caption{\label{fig:round22a}}%
\end{subfigure}\hfill
\begin{subfigure}[c]{.3\textwidth}
  \centering
  \includegraphics[width=\linewidth]{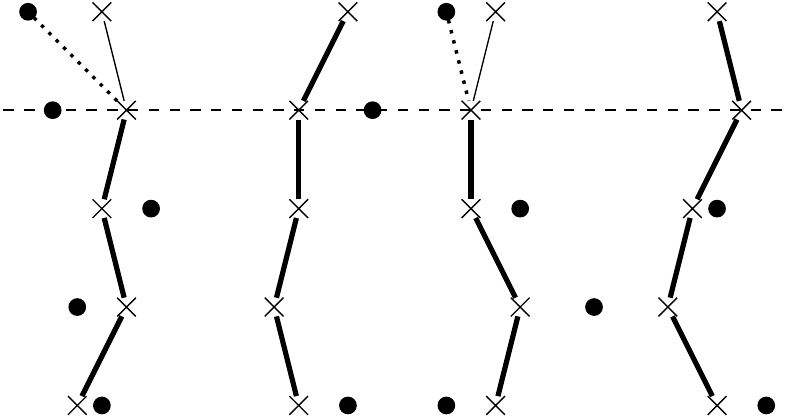}
  \caption{\label{fig:round22b}}%
\end{subfigure}\hfill
\begin{subfigure}[c]{.3\textwidth}
  \centering
  \includegraphics[width=\linewidth]{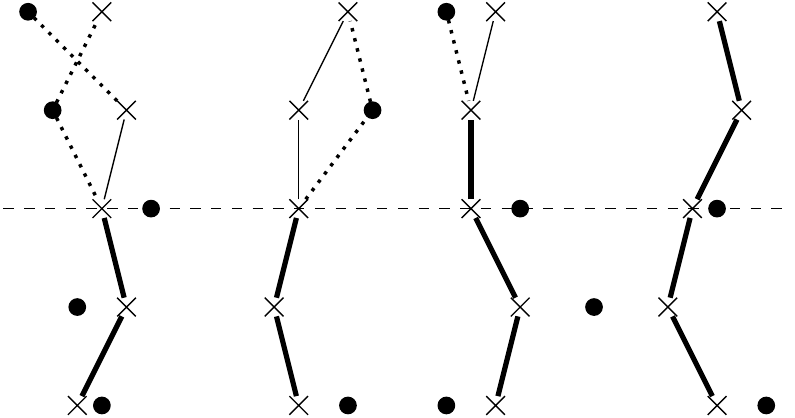}
  \caption{\label{fig:round22c}}%
\end{subfigure}
\caption{%
An example of the inductive construction of the multisets of paths ${\cal Q}_i$,
for $i=0$ (Figure \ref{fig:round22a}), $i=1$ (Figure \ref{fig:round22b}), and
$i=2$ (Figure \ref{fig:round22c}). Dotted lines show where a trajectory has been
rounded to a net point. Thin and thick solid lines indicate where one or two
trajectories are coincident to an optimal trajectory, respectively. Initially
(Figure \ref{fig:round22a}), ${\cal Q}_0$ consists of $2k$ trajectories
$\sigma^1_j=\sigma^2_j=\tau_j$, for the trajectories of some optimal solution
$\tau_1, \ldots, \tau_k$. At step $i$, for any $j\in[k]$ such that $\tau_j(i)$
is within a distance of $r$ from some net point, $c$, we obtain ${\cal Q}_i$ by
replacing $\tau_j(i)$ with $c$ in either $\sigma^1_j$ or $\sigma^2_j$, depending
on the parity of $i$.\label{fig:2k2r}
}
\end{figure}

\begin{proof}
Fix a \triclus/ $\mathcal{C}=\{\tau_i\}_{i=1}^k$. We inductively define a
sequence ${\cal Q}_0,\ldots,{\cal Q}_t$, where for each $i\in \{0,\ldots,t\}$,
${\cal Q}_i$ is a multiset of paths in $G_{r+\delta}(P)$. We set ${\cal
Q}_0=\{\sigma^1_1,\sigma^2_1,\ldots,\sigma^1_k,\sigma^2_k\}$, where for each
$j\in [k]$, we have $\sigma_j^1=\sigma_j^2=\tau_j$. Next, we inductively define
${\cal Q}_i$, for some $i\in \{1,\ldots,t\}$. Starting with ${\cal Q}_i={\cal
Q}_{i-1}$, we proceed to modify ${\cal Q}_i$.
By induction, it follows that the paths $\sigma^1_j$,  $\sigma^2_j$, and
$\tau_j$ share the same suffix at levels $i,\ldots,t$. Thus, $\tau_j(i)\in
\sigma^1_j$ and $\tau_j(i)\in \sigma^2_j$. Now, for the modification, we
consider each $c\in C(i)$, and proceed as follows (see Figure~\ref{fig:2k2r} for
an illustration). Since $\mathcal{C}$ is a valid \triclus/, it follows from
Lemma~\ref{lem:injective} that there exists an injective map $\pi_i$ from $C(i)$
to the set $\tau_1(i),\ldots,\tau_k(i)$ so that $\pi_i(c)=\tau_j(i)$ for some
$j\in [k]$ and $d(\tau_j(i), c) \leq r$. We consider the following two cases:
\begin{description}[topsep=0pt,itemsep=0pt]
\item{Case 1:} If $i$ is odd and $\tau_j(i)=\pi_i(c)$ for some $c\in C(i)$, then
we modify $\sigma^1_j$ by replacing the vertex $\tau_j(i)$ with $c$.
\item{Case 2:} If $i$ is even and $\tau_j(i)=\pi_i(c)$ for some $c\in C(i)$,
then we modify $\sigma^2_j$ by replacing the vertex $\tau_j(i)$ with $c$.
\end{description}
We next argue that the result is indeed a path in $G_{r+\delta}(P)$. Suppose
that in the above step, we modify the path $\sigma^\ell_j$, for some $\ell\in
\{1,2\}$ so that $\sigma^\ell_j(i)=c$. It follows by induction on $i$ that the
path $\sigma^\ell_j$ was not modified when constructing ${\cal Q}_{i-1}$; thus
$\sigma^\ell_j(i-1) = \tau_j(i-1)$. Since $\delta(\tau_j)\leq r$, it follows by
the triangle inequality that
$
  d(\sigma^\ell_j(i-1), \sigma^\ell_j(i)) = d(\tau_j(i-1), c)
    \leq d(\tau_j(i-1), \tau_j(i)) + d(\tau_j(i), c)
    \leq \delta+r.
$
It follows that $\delta(\sigma^{\ell}_j)\leq r + \delta$, which implies that
each element of ${\cal Q}_i$ is indeed a path in $G_{r+\delta}(P)$. This
completes the inductive definition of the multisets ${\cal Q}_0,\ldots,{\cal
Q}_t$.
It is immediate by induction that for each $i\in [t]$, for each $c\in C(i)$,
there exist some path $\sigma\in {\cal Q}_t$ that visits $c$.
We next transform the collection ${\cal Q}_t$ into a flow $F$ in $N_{r +
\delta}(P,C)$. For each path $\sigma\in {\cal Q}_t$, we obtain a path in the
network $N_{r+\delta}(P,C)$ starting from the source $s$, then replacing for
each $i\in [t]$, each $c\in C(i)\cap \sigma$ by the edge $(\tail(v), \head(v))$,
for $v = (i,c)$, then terminating at the sink $s'$; we route a unit of flow
along the resulting path. Since for each $i\in [t]$, there exists some path in
${\cal Q}_t$ visiting each $c\in C(i)$, it follows that all lower-bound
constraints in $N_{r+\delta}(P,C)$ are satisfied by $F$. Since ${\cal Q}_t$
contains $2k$ paths, it follows that the value of the resulting flow is $2k$, as
required.
\end{proof}

We are now ready to prove Theorem~\ref{thm:BICRIT-approx}.

\begin{proof}[Proof of Theorem~\ref{thm:BICRIT-approx}]
For each $i\in [t]$, compute a $2r$-net $C(i)$ of $P(i)$, and construct the flow
network $N_{r+\delta}(P,C)$. Compute a minimum flow $F$ in $N_{r+\delta}(P,C)$
satisfying all lower-bound constraints. If $N_{r+\delta}(P,C)$ is infeasible
(i.e.~if there is no flow satisfying all lower bound constraints), or if the the
value of the minimum flow in $N_{r+\delta}(P,C)$ is greater than $2k$, it
follows by Lemma \ref{lem:flow_2r} that $P$ does not admit a \triclus/. Thus, in
this case the algorithm terminates. Otherwise, we compute a minimum flow in
$N_{r+\delta}(P,C)$. Since all capacities and lower-bound constraints in
$N_{r+\delta}(P,C)$ are integers, it follows that $F$ can be taken to be
integral. We decompose $F$ into a collection of at most $2k$ paths, each
carrying a unit of flow. Arguing as in Lemma \ref{lem:DKC} we have that the
restriction of these paths on $G_{r+\delta}(P)$ is a set of trajectories that
induces a valid \triclusp{2k}{2r}{r+\delta} of $P$. This provides a
\triappp{2}{2}{1+\eps} where $\eps = r / \delta$. Finally, the running time is
easily seen as $O(n^3)$ by the same argument that appears in
Lemma~\ref{lem:DKC}, concluding the proof.
\end{proof}

\subsection{Approximation algorithm for temporal median clustering}
\label{sec:tkmapprox}
\begin{figure}[h]%
\centering
\begin{subfigure}{0.35\linewidth}
  \includegraphics[width=\linewidth]{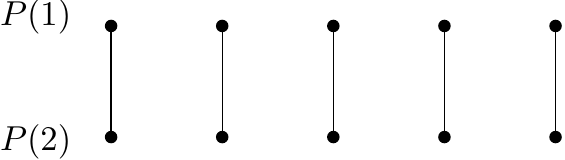}
  \caption{The clustering $\mathcal{C}^*$.}
  \label{fig:local-search-opt}%
\end{subfigure}
\hspace{1cm}
\begin{subfigure}{0.35\linewidth}%
  \includegraphics[width=\linewidth]{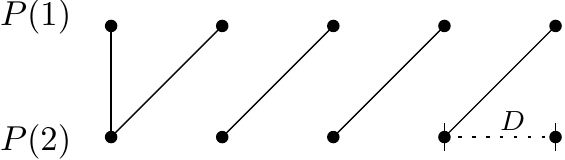}
  \caption{The clustering $\mathcal{C}$.}
  \label{fig:local-search-stuck}%
\end{subfigure}
\caption{%
    An example demonstrating that local search fails. Consider a
    temporal-sampling $P$ of length $2$ where $P(1) = P(2)$ consists of a
    sequence of $5$ points where successive points are separated by $D$.
    (\ref{fig:local-search-opt}) A temporal $(5,r,\delta)$-median-clustering,
    $\mathcal{C}^*$, for any $r, \delta \in \mathbb{R}_{\geq 0}$ with
    $\mediancost{(\mathcal{C}^*)} = 0$.
    (\ref{fig:local-search-stuck}) A temporal
    $(5,D,\delta)$-median-clustering, $\mathcal{C}$, for any $D \leq
    \delta < 2 D$. Note that swapping any trajectory in $\mathcal{C}$ with
    one in $\mathcal{T}_\delta(P)$ is non-improving. The clustering
    $\mathcal{C}$ is therefore a local minimum of local search, yet the ratio
    $\mediancost{(\mathcal{C})}/\mediancost{(\mathcal{C}^*)}$ remains unbounded.
    \label{fig:local-search-fails}
    }%
\end{figure}

In this section we consider variants of \TC/ which evaluate the spatial cost of
clustering by taking the level-wise maximum of discrete $k$-median and discrete
$k$-means objectives. A natural question is whether or not the problem admits a
$O(1)$-approximation via local search, as in static case
\cite{arya2004local,korupolu2000analysis}. In
Figure~\ref{fig:local-search-fails} we show that the local search approach
fails, even on temporal samplings of length two. Instead, the result is obtained
by iteratively selecting a trajectory which most improves a certain potential
function. The result in this section is presented for the \TKMEDC/ problem,
and follows by submodularity and monotonicity of the potential function. These
properties remain if $d(p, {\cal C}(i))$ is replaced with the $d(p, {\cal
C}(i))^2$, and thus holds identically for \TKMEANS/.

We now present an approximation algorithm for the \TKMEDC/ problem. Let
$\mathcal{I}=(M,P,k,r,\delta)$ be an input to the problem, where $P$ is a
temporal-sampling of length $t$. Let $n$ denote the size of the $P$. Let also
$\Delta$ denote the spread of $M=(X,d)$. That is,
$
\Delta = \frac{\diam(M)}{\inf_{p,q \in X} \{d(p,q): d(p,q) > 0\}}.
$
Since we only consider finite metric spaces, and since the single point case is
trivial, w.l.o.g.~we may assume that the diameter of $M$ is $\Delta$ and
minimum interpoint distance in $M$ is $1$. For a set of trajectories ${\cal C}$
we define
$
\cost(i; {\cal C}) = \sum_{p\in P(i)} d(p, {\cal C}(i)).
$
We also define
$
W({\cal C}) = \sum_{i=1}^t \max\{0, \cost(i; {\cal C})-r\}.
$
Intuitively, the quantity $W({\cal C})$ measures how far the solution ${\cal C}$
is from the optimum; in particular, if $W({\cal C})=0$ then the spatial cost is
within the desired bound.

\begin{lemma}\label{lem:tasos:submodular}
The set function $-W$ is submodular.
\end{lemma}
\begin{proof}
Since the sum of submodular functions is submodular, it is enough to show that
$-\max\{0, \cost(i; {\cal C})-r\}=\min\{0,-\cost(i;{\cal C})+r\}$ is submodular.
Thus it suffices to show that $-\cost(i;{\cal C})$ is submodular, and thus it
suffices to show that $-d(p,{\cal C}(i))$, for all $p\in P(i)$, which is
immediate since $d(p,{\cal C}(i)) = \min_{\tau\in {\cal C}} d(p,\tau(i))$.
\end{proof}

\textbf{Algorithm.}
Our goal is to compute some set of trajectories ${\cal C}$ such that $W({\cal
C})$ is sufficiently small, while minimizing $|{\cal C}|$. The algorithm
consists of the following steps:
\begin{description}[topsep=0pt,itemsep=0pt]
\item{\textbf{Step 1.}}
Let ${\cal C}_0$ be a set containing a single arbitrary trajectory.
\item{\textbf{Step 2.}}
For any $i\in [L]$,
let $\tau_i$ be a minimizer of $W({\cal C}_{i-1}\cup \{\tau_i\})$.
Set ${\cal C}_i = {\cal C}_{i-1}\cup \{\tau_i\}.$
\item{\textbf{Step 3.}}
Return ${\cal C}_L$.
\end{description}
The parameter $L>0$ will be determined later.

The following Lemma bounds the running time of Step 2.

\begin{lemma}\label{lem:tasos:time}
Given a clustering ${\cal C}$, we can find $\tau$
minimizing $W({\cal C}\cup \{\tau\})$, in time $\poly(|{\cal C}|,n)$.
\end{lemma}

\begin{proof}
This can be done via dynamic programming. The proof is essentially the same as
in Lemma~\ref{lem:find_path} and is thus omitted.
\end{proof}

We next show that for some value of the parameter $L$, the algorithm computes a
solution with low cost. To that end, we show in the next Lemma that at each
iteration of the main loop the quantity $W({\cal C}_i)$ decreases by a
significant amount.

\begin{lemma}\label{lem:tasos:improvement}
If $\mathcal{I}$ admits a temporal $(k,r,\delta)$-median-clustering, then for
any $i\in \{1,\ldots,L\}$, there exists some feasible trajectory $\sigma_i$ such
that $W({\cal C}_{i-1}\cup \{\sigma_i\}) \leq (1-1/k) \cdot W({\cal C}_{i-1})$.
\end{lemma}
\begin{proof}
Let $\mathcal{C}^*=\{\tau^*_1,\ldots,\tau_{k'}^*\}$ be a set of at most $k$
trajectories that yields a $(k,r,\delta)$-median temporal clustering.
W.l.o.g.~we may assume that $k' = k$. Let $K_0=W({\cal C}_{i-1})$, and for any
$j\in [k]$, let
$
K_j = W({\cal C}_{i-1} \cup \{\tau^*_1,\ldots,\tau^*_j\}).
$
Since $\mathcal{C}^*$ is a $(k,r,\delta)$-median temporal clustering, it follows
that $W({\cal C}_{i-1}) = K_0 \geq K_1 \geq \ldots \geq K_k = 0.$
For any $j\in [k]$, we also define $K'_j = W({\cal C}_{i-1} \cup \{\tau^*_j\})$.
By Lemma \ref{lem:tasos:submodular} we have that for all $j\in [k]$, $W({\cal
C}_{i-1}) - W({\cal C}_{i-1} \cup \{\tau^*_j\}) \geq W({\cal C}_{i-1} \cup
\{\tau^*_1,\ldots,\tau^*_{j-1}\}) - W({\cal C}_{i-1} \cup
\{\tau^*_1,\ldots,\tau^*_j\})$. That is, $K_0-K'_j \geq K_{j-1} - K_j$.
Let $\ell = \argmax_{j\in [k]} \{K_0-K'_j\}.$
It follows that
$
  K_0 - K'_{\ell} = \max_{j\in [k]} \{K_0-K_j'\} \geq \max_{j\in [k]}\{K_{j-1}-K_j\}
    \geq \frac{1}{k}\sum_{j=1}^k (K_{j-1}-K_j) = (K_0-K_k)/k = K_0/k.
$
Let $\sigma_i = \tau^*_{\ell}$. It immediately follows that $W({\cal C}_{i-1}\cup
\{\sigma_i\}) = K'_{\ell} \leq (1-1/k) \cdot K_0 = (1-1/k) \cdot W({\cal
C}_{i-1})$, concluding the proof.
\end{proof}

We are now ready to prove Theorem~\ref{thm:kmlogeps1}.

\begin{proof}[Proof of Theorem~\ref{thm:kmlogeps1}]
We first note that if $r=0$, then a solution with $k$ trajectories can be
computed, if one exists, as follows: Since $r=0$, it follows that every level of
$P$ has at most $k$ points. We construct the flow network instance
$N_{\delta}(P,P)$, as in Section \ref{sec:k-center}. It is immediate that the
flow instance is feasible iff there exists a solution with $k$ trajectories.
For the remainder of the proof we may thus assume that $r>0$. Since the minimum
distance in $M$ is 1, it follows that $r\geq 1$. In a generic step $1 \leq i
\leq L$, let $\tau_i$ denote the trajectory returned by the dynamic program of
Lemma~\ref{lem:tasos:time}, which minimizes $W({\cal C}\cup \{\tau_i\})$. By Lemma \ref{lem:tasos:improvement} we have that if $\mathcal{I}$
admits a temporal $(k,r,\delta)$-median-clustering, then there exists some trajectory
$\sigma_i$ such that $W({\cal C}_{i-1}\cup \{\sigma_i\}) \leq
(1-1/k) \cdot W({\cal C}_{i-1})$. Thus
$
W({\cal C}_i) = W({\cal C}_{i-1}\cup \{\tau_i\})
 \leq W({\cal C}_{i-1} \cup \{\sigma_i\})
 \leq (1-1/k) \cdot W({\cal C}_{i-1})
 \leq (1-1/k)^i \cdot W({\cal C}_0)
$.
Since the diameter of $M$ is $\Delta$, we get $W({\cal C}_0) \leq \Delta \sum_{i
\in [t]} |P(i)| = \Delta n$. Setting
$L=k\ln(n\Delta/\eps)=O(k\log(n\Delta/\eps))$, we obtain $W({\cal C}_L) \leq
(1-1/k)^L n \Delta \leq \eps \leq \eps r.$ Thus
$
\max_{i\in [t]} \max\{0,\cost(i;{\cal C}_L)-r\} \leq \sum_{i=1}^{t} \max\{0,\cost(i;{\cal C}_L)-r\} \leq \eps r,
$
which implies $\mediancost({\cal C}_L) = \max_{i\in [t]} \cost(i;
{\cal C}) \leq (1+\eps) r$. It follows that either the computed solution ${\cal
C}_L$ is a $(L,1+\eps,1)$-approximation, or $\mathcal{I}$ does not admit a
temporal $(k,r,\delta)$-median-clustering, as required.
Finally, the bound on the running time follows by the fact that we perform $L$
iterations of the main loop; the running time of each iteration is bounded by
Lemma \ref{lem:tasos:time}.
\end{proof}

\section{Inapproximability}
In section~\ref{sec:nphardsmalldelta} we show that temporal-clustering with an
exact number of clusters is \NP-hard to obtain a
$(1,\poly(n),\poly(n))$-approximation (Theorem~\ref{thm:1polypolyhard}),
complementing Theorem~\ref{thm:DKC}. In section~\ref{sec:setchard} we argue
that the $(\ln n, 1, 1)$-approximation given in Theorem~\ref{thm:DRDS-approx} is
best possible by observing that $((1-\eps)\ln n, 2-\eps',\cdot)$-approximation
is \NP-hard (Theorem~\ref{thm:hardassetcover}), though the construction involves
a somewhat unnatural metric space. In section~\ref{sec:inapprox2} we show that
$(1.005,2-\eps,\poly(n))$-approximation is \NP-hard even for points sampled from
$2$-dimensional Euclidean space (Theorem~\ref{thm:ccpolyhard}). In
section~\ref{sec:1polypolyhardkm} (Theorem~\ref{thm:1polypolyhardkm}) and
section~\ref{sec:ccpolyhardkm} (Theorem~\ref{thm:ccpolyhardkm}), we adapt the
hardness results for $(1,\poly(n),\poly(n))$-approximation, and
$(1.005,2-\eps,\poly(n))$-approximation to the \TKMED/ setting. We observe that
these constructions involve clusterings which use only a constant number of
points per cluster, thus the same constructions suffice to prove hardness of
\TKMEANSC/ with only slight modification of the distances.

\subsection{Inapproximability with exact number of clusters}
\label{sec:nphardsmalldelta}
We complement Theorem~\ref{thm:DKC} by showing that it is \NP-hard to obtain a
$(1,\poly(n),\poly(n))$-approximation. Further, this inapproximability result
holds even for a temporal sampling in $\mathbb{R}^2$. Let $P$ be such a sample
consisting of $n$ points. We show that given any fixed $\eps> 0$ and $s \in
[0,1]$ there exists universal constants $c_r$, $c_\delta$ such that the
\triappp{1}{c_rn^{s(1-\eps)}}{c_\delta n^{(1-s)(1-\eps)}} problem is $\NP$-hard.
We describe a polynomial-time reduction from instances of \THREESAT/ to
temporal-samplings over $\mathbb{R}^2$. In particular, given any positive real
numbers $\eps$, $s \leq 1$, $r_0$, and $\delta_0 < r_0 \sqrt{3}/4$, and some
$l$-variable instance ${\cal S}$ of \THREESAT/, we construct a temporal sampling
$P$ such that the following conditions hold:
\begin{enumerate}[topsep=0pt,itemsep=0pt]
  \item $P$ admits a \triclusp{l}{r_0}{\delta_0} if ${\cal S}$ is satisfiable.
  \item $P$ does not admit a \triclusp{l}{c_rn^{s(1-\eps)}r_0}{c_\delta n^{(1-s)(1-\eps)}\delta_0} otherwise.
\end{enumerate}

\begin{figure}
\centering
\begin{subfigure}[b]{.47\textwidth}
\includegraphics[width=\textwidth]{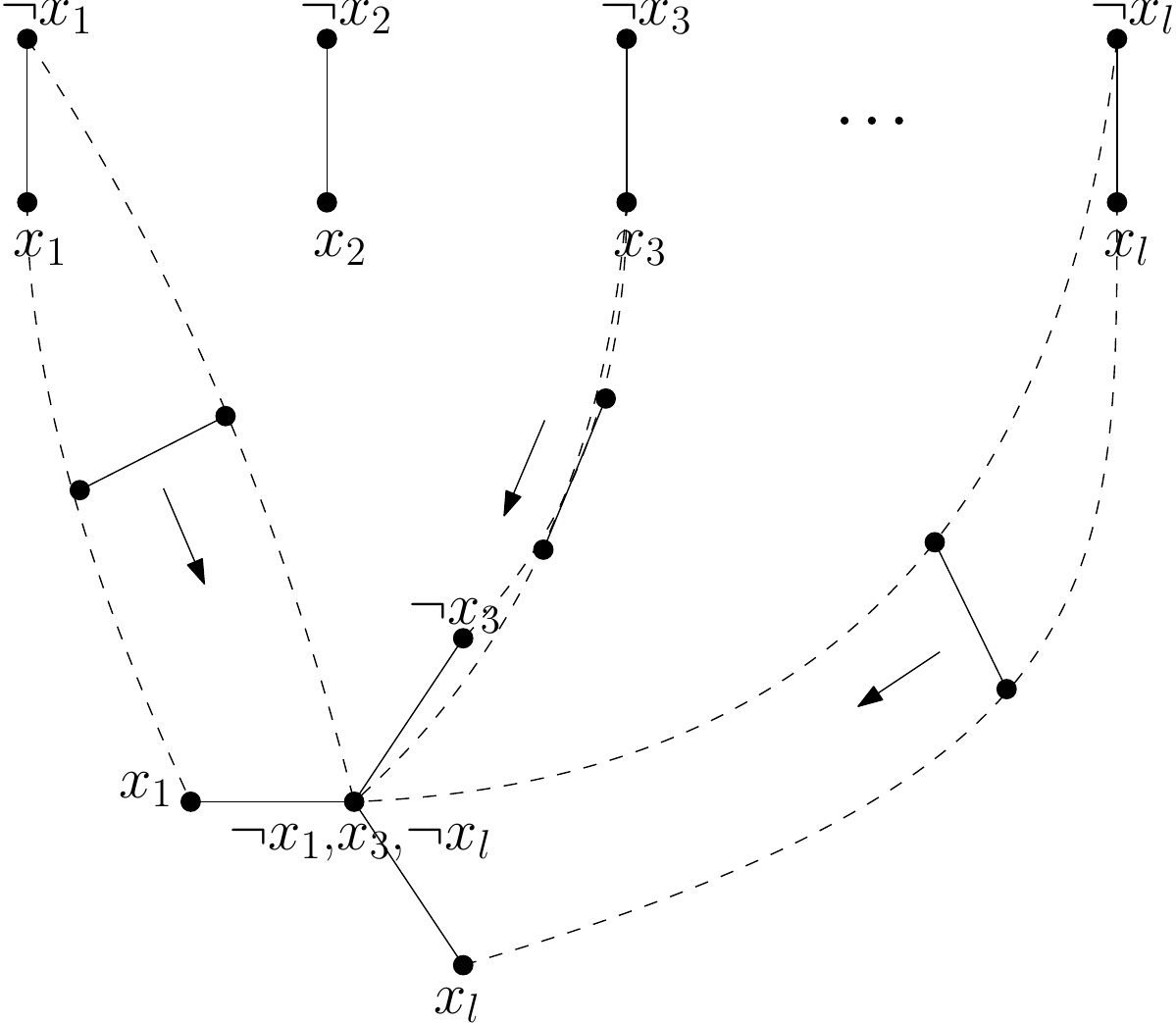}
\caption{Assembling the clause $\neg x_1 \vee x_3 \vee \neg x_l$.\label{fig:assembly}}
\end{subfigure}\qquad
\begin{subfigure}[b]{.47\textwidth}
\includegraphics[width=\linewidth]{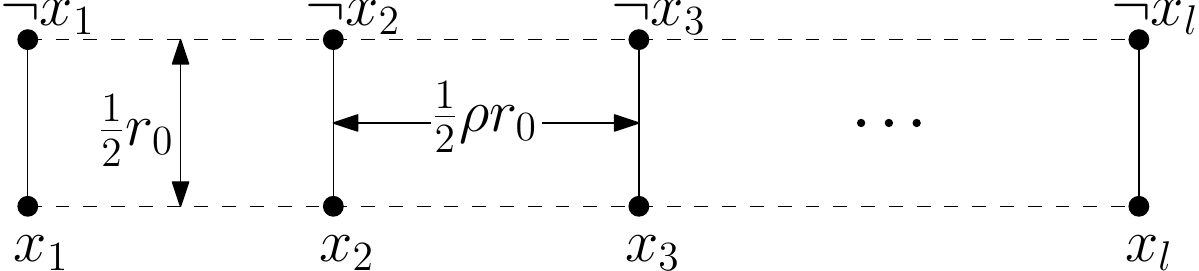}
\caption{Initial position for some $\rho \geq 1$.\label{fig:initial}}%
\vspace{2.75ex}
\includegraphics[width=\linewidth]{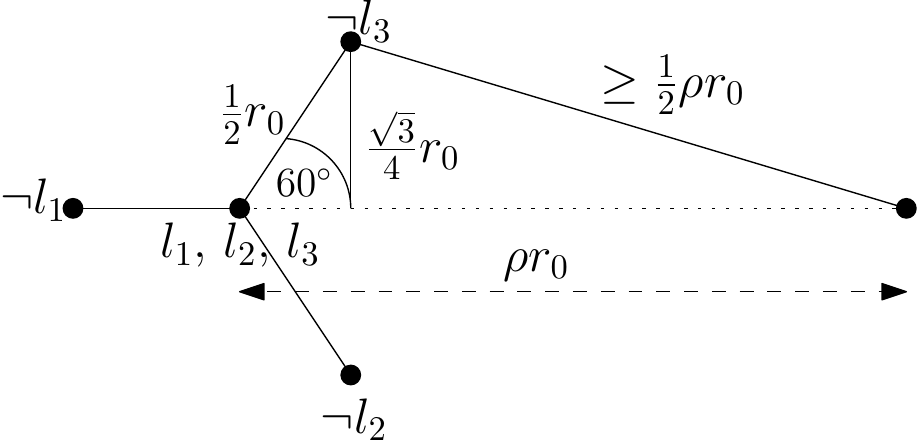}
\caption{A fully assembled clause gadget.\label{fig:clause}}%
\end{subfigure}
\caption{Reference diagrams for gadget construction.}
\end{figure}

Suppose we are given an instance ${\cal S}$ of \THREESAT/ with $l$ variables and
$m$ clauses. W.l.o.g., we assume that every clause
contains no repeated variables. We now describe how to produce the corresponding
temporal-sampling $P$:

\textbf{Variable gadgets.}
For each variable $x_i$ of ${\cal S}$ we introduce a pair of points in each
level of $P$. We denote these points by $x_i$ and $\neg x_i$ and the pair
$\{x_i, \neg x_i\}$ by $v_i$. We will sometimes refer to these points as
literals. Initially, we lay out the variable gadgets in the plane such that
$d(x_i, \neg x_i) = \frac{1}{2}r_0$ for all $1 \leq i \leq l$, and $\rho r_0/2
\leq d(v_i, v_j) \leq l \rho r_0/2$ for all $1 \leq i < j \leq l$. Here, $\rho =
\rho(l) \geq 1$ denotes a yet to be determined function of the number of
variables (see Figure~\ref{fig:initial}). We refer to this configuration as
\emph{initial position}.

\textbf{Clause gadgets.}
Order the clauses of ${\cal S}$ arbitrarily as $c_1, \ldots, c_m$. For each
clause we build a series of levels for the temporal-sampling where the variable
gadgets of that clause appear to undergo rigid motion see
(Figure~\ref{fig:assembly}). By \emph{motion} we mean that points in any pair of
subsequent levels which correspond to the same literal are within a bounded
distance of each other. Further, we say it is \emph{rigid} because in every
level we maintain the distance of $\frac{1}{2}r_0$ between literals of the same
variable gadget. Note that we may enforce a consistent assignment of a center to
a literal between consecutive levels by ensuring that the distance between a
literal and its copy in the next level is at most $\delta_0$, and that the
distance between points corresponding to distinct literals exceeds $\delta_0$.
We describe this motion in three phases:
%At the start of the first phase and at
%the end of the last phase the points are in the initial position.
%This ensures that the motions are \emph{composable}
%in the sense that assignments of centers to literals at the end of a motion for
%one clause remain the same at the beginning of the next.
%
\begin{description}[topsep=0pt,itemsep=0pt]
\item{\textbf{Phase 1: Assembly.}} All points start in the initial position (see
Figure~\ref{fig:initial}). The variable gadgets which are used in the specified
clause are brought together one by one under rigid motion (see
Figure~\ref{fig:assembly}). This motion brings the ends of the variable gadgets
which correspond to the literals that appear in the clause to a single common
point. The unused literals appear on a circle of radius $\frac{1}{2}r_0$ about
the common point, arranged such that they form an equilateral triangle (see
Figure~\ref{fig:clause}). This phase takes $O(l \rho r_0 / \delta_0)$ steps per
clause.
\item{\textbf{Phase 2: Satisfiability check.}} An extra point is introduced at
the location where the variable gadgets meet. This point then undergoes motion
directly away from one of the unused literals for a distance of $\rho r_0$,
before reversing course and returning to the common point at the center of the
gadget (see Figure~\ref{fig:clause}). It subsequently disappears. This phase
takes $O(\rho r_0 / \delta_0)$ steps per clause.
\item{\textbf{Phase 3: Disassembly.}} The motion of \emph{Phase 1} is reversed
and the variable gadgets are returned to initial position in $O(l \rho r_0 /
\delta_0)$ steps per clause.
\end{description}
\textbf{Analysis.}
Let $P$ denote the temporal-sampling of $\mathbb{R}^2$ given by the above
construction, and let $\opt{r}$ denote the smallest value of $r$ such that $P$
admits a \triclusp{l}{r}{\delta_0}.

\begin{lemma}\label{lem:sat}
If ${\cal S}$ is satisfiable then $\opt{r} \leq r_0$.
\end{lemma}

\begin{proof}
Since ${\cal S}$ is satisfiable there exists a satisfying assignment. We now
exhibit a solution of cost at most $r_0$. For each of the $l$ variable gadgets,
we set one of the two points to be a center in the initial configuration. If
$x_i$ is \TRUE/ in the satisfying assignment then the point which corresponds to
the literal $x_i$ is selected as a center. Otherwise, $\neg x_i$ is selected.
Note that the construction of the gadget ensures that the same choice of
literals can be maintained throughout the entire motion. We maintain these
choices during \emph{Phase 1} and \emph{Phase 3} where the only points which
appear are from variable gadgets. Since at least one side of each variable
gadget is covered and the distance between the points of these gadgets are
$\frac{1}{2}r_0$ at all times, the covering cost of each level in these phases
is $\frac{1}{2}r_0$. By satisfiability there is at least one center at the
common point where the clause literals meet. We take one such center and
assign it to the extra point. Throughout \emph{Phase 2} the extra point is
covered and the clause gadget retains at least one center (somewhere) at a
coverage cost of at most $r_0$.
\end{proof}

\begin{lemma}\label{lem:unsat}
If ${\cal S}$ is not satisfiable then $\opt{r} \geq \frac{1}{2} \rho r_0$.
\end{lemma}

\begin{proof}
We will show that a satisfying assignment can be inferred from a clustering
${\cal C}$ with cost below $\frac{1}{2} \rho r_0$.
First, we argue that the variable gadgets are consistent in initial position.
Since $\rad({\cal C}) < \frac{1}{2}\rho r_0$, it follows that whenever the point
configuration is in initial position, every variable gadget has exactly one
center. Thus we do not simultaneously select a literal and its negation, as
otherwise it follows by the pigeonhole principle that at least one variable
gadget is uncovered and $\rad({\cal C}) \geq \rho r_0$.
Next, we argue that these choices must remain consistent within a clause gadget
$c=(l_1 \vee l_2 \vee l_3)$. The only opportunity for a trajectory to change
literals is when two literals are within a distance of $\delta_0$, and the only
literals in the gadget which pass within $\delta_0$ are $l_1$, $l_2$, and $l_3$.
To see that inconsistency is expensive, assume that $l_1$ is a center in the
first level of the gadget but not in the last, then the variable gadget which
contains literal $l_1$ is uncovered in the last level for the clause, implying
$\rad({\cal C}) \geq \rho r_0$. Moreover, since $\delta_0 < r_0$ the choice of
cluster centers at the end of clause gadget $c_i$ must be the same as at the
beginning of $c_{i+1}$ for $1 \leq i < t$.
We now argue that at least one of the literals for a clause is a center. Suppose
none of them are a center, then the extra point from \emph{Phase 2} never
coincides with a center as $\delta_0 < r_0 \sqrt{3}/4$ and the closest
possible location for a center to any point along its motion is at a distance of
$r_0 \sqrt{3}/4$ away. At its maximum displacement the extra point is at a
distance of $\rho r_0$ from where the literals meet, which is at least
$\frac{1}{2} \rho r_0$ from any other center. Thus, $\rad({\cal C}) \geq
\frac{1}{2} \rho r_0$. We produce a satisfying assignment by setting true those
literals which are assigned centers in ${\cal C}$.
\end{proof}

\begin{remark}\label{rem:scale}
We remark that Lemma~\ref{lem:sat} and Lemma~\ref{lem:unsat} continue to hold
even for $\delta = \beta \delta_0$ for any $1 \leq
\beta<(\sqrt{3}/4)r_0/\delta_0$. Essentially, there are two important distance
scales given by $\delta_0$ and $r_0$. Points in successive levels closer than
$\beta \delta_0$ can be assumed to be $\delta_0$ and vice versa.
%In other words, let $P$ denote the
%temporal-sampling of $\mathbb{R}^2$ given by the above construction, and let
%$\opt{r'}$ denote the smallest value of $r$ such that $P$ admits a
%\triclusp{l}{r}{\beta \delta_0} for any $1 \leq \beta<(\sqrt{3}/4)r_0/\delta_0$.
%Then $\opt{r'} = \opt{r}$.
%
\end{remark}

\begin{lemma}\label{lem:convtime}
Given an instance ${\cal S}$ of \THREESAT/ with $l$ variables and $m$ clauses.
It is possible to construct the above temporal-sampling of size $n \in O(\rho
r_0 l^2 m / \delta_0)$ in $O(\rho r_0 l^2 m / \delta_0)$-time.
%In general we think of $\rho$, $r_0$, and $\delta_0$ as functions of the input.
%In particular, the running time is polynomial in the size of ${\cal S}$ and in $\rho r_0 / \delta_0$.
%
\end{lemma}

\begin{proof}
This is immediate by construction, as each clause gadget consists of
$O(\rho r_0 l / \delta_0)$ levels of size $O(l)$ and there are $m$ clauses.
\end{proof}

We are now ready to prove Theorem~\ref{thm:1polypolyhard}.

\begin{proof}[Proof of Theorem~\ref{thm:1polypolyhard}]
Let $c > 5(1/\eps - 1)$. Let ${\cal S}$ be an instance of \THREESAT/ with $l$
variables and $m$ clauses. We invoke Lemma~\ref{lem:convtime} with
$\rho(l)=l^{s\cdot c}$, $r_0(l)/\delta_0(l) = l^{(1-s)\cdot c}$ and to yield, in
polynomial-time, a temporal-sampling $P$ over $\mathbb{R}^2$ of size $n \in
O(l^{c+5})$. Note here we have used the fact that $m \in O(l^3)$. Suppose the
existence of some polynomial-time \triappp{1}{\alpha(n)}{\beta(n)} where
$\alpha(n) = c_r n^{s\cdot(1-\eps)}$, and $\beta(n) = c_\delta
n^{(1-s)(1-\eps)}$. There exists a universal constant $c_1$ (where $c_1 > 1$
by construction) such that $c_1 n^{(1-s)\cdot(1-\eps)} \leq (\sqrt{3}/4)
r_0/\delta_0 = (\sqrt{3}/4) l^{(1-s)\cdot c}$. Thus we let $1 \leq c_\delta <
c_1$, Remark~\ref{rem:scale} indicates that Lemma~\ref{lem:sat} and
Lemma~\ref{lem:unsat} still apply. We run this approximation on $P$ with $k=l$,
$r=r_0$, and $\delta=\delta_0$. In polynomial-time the algorithm either produces
output or fails. If the algorithm fails it follows by
Definition~\ref{def:APPROXTRICLUS} that $P$ does not admit a
\triclusp{l}{r_0}{\delta_0} and thus ${\cal S}$ is not satisfiable by
Lemma~\ref{lem:sat}. Otherwise, the algorithm produces some clustering ${\cal
C}$ of radius $\rad(\mathcal{C}) \leq \alpha(n) r_0$. Lemma~\ref{lem:unsat}
ensures that if $\mathcal{S}$ is not satisfiable then $\rad(\mathcal{C})
\geq\frac{1}{2}\rho r_0 = \frac{1}{2} l^{s\cdot c} r_0 \geq c_0 n^{s c/(c+5)}
r_0 \geq c_0 n^{s\cdot(1-\eps)} r_0$ for some universal constant $c_0 > 0$. It
follows that if $c_r \in (0, c_0)$ the algorithm produces output if and only if
${\cal S}$ is satisfiable, giving a polynomial-time test for satisfiability.
\end{proof}

\subsection{\NP-hardness of \triappp{(1-\eps)\ln(n)}{2-\eps'}{1}}\label{sec:setchard}
We now argue that the \triappp{\ln(n)}{1}{1} is tight. The \RDS/
problem for a metric space $(X, d)$ is to find a smallest set of points $C$ such
that every point in $X$ is at most a distance $r$ from a point in $C$.
It is known that \SETC/ reduces to \RDS/ in polynomial-time \cite{Eisenstat14}.
For completeness we now give such a reduction:

\begin{folklore}
\label{lem:setc-to-rds}
\SETC/ reduces to \RDS/ in polynomial-time. Moreover, any
$\alpha(n)$-approximate solver for \RDS/ yields an $\alpha(n)$-approximation for
\SETC/.
\end{folklore}

\begin{figure}
\begin{center}
\includegraphics[width=0.4\textwidth]{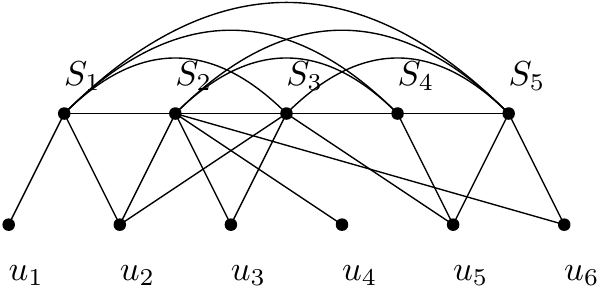}
\end{center}
\caption{\label{fig:setc-to-rds}
The \RDS/ instance corresponding to the \SETC/ instance $(U,\mathcal{S})$ with
$U=\{u_1,\ldots,u_6\}$ and $S=\{S_1, \ldots, S_5\}$, where $S_1=\{u_1, u_2\}$,
$S_2=\{u_2, u_3, u_4, u_6\}$, $S_3=\{u_2, u_3, u_5\}$, $S_4=\{u_5\}$,
$S_5=\{u_5, u_6\}$. Edges are shown between points at distance of $1$. All other
pairs of distinct points have a distance of $2$.
}
\end{figure}
\begin{proof}
Let $(U, \mathcal{S})$ be an instance of \SETC/, where
$\mathcal{S}=\{S_1,\ldots,S_n\}$. We define a metric space $M=(X,d)$ where $X$
contains points corresponding to the elements of $U$, and $\mathcal{S}$, and
where $d$ satisfies the following:
\begin{enumerate}
\item $d(S_i,S_j)=1$ for all $S_i, S_j \in \mathcal{S}$, $i \neq j$.
\item $d(u,S_i)=1$ for any $S_i \in \mathcal{S}$, $u$ such that $u \in S_i$.
\item Otherwise, $d(u,v)=2$ for any remaining pair of distinct points $u,v\in X$.
\end{enumerate}
These constraints on $d$, together with the requirement that it be a metric on
$M$, completely determine $d$. (See Figure~\ref{fig:setc-to-rds} for an
illustration.)

Let $S$ be a feasible solution to the \SETC/ instance. Note that $S$ induces a
{\textsc {$1$-Dominating Set}} solution of size $|S|$ by taking the set of
points in $X$ which correspond to the points in $S$. On the other hand, any
feasible solution of {\textsc {$1$-Dominating Set}} can be converted into a
feasible solution of \SETC/ in linear time without increasing its size. To see
this fix a feasible solution $R' \subset X$, and let $u \in R' \cap U$. Note
that a $1$-ball of $u$ in $M$ consists only of $u$ and points for elements of
$\mathcal{S}$ which cover it. Instead we can cover $u$ by any other point in its
$1$-ball, $S_i$. Since $\ball(u,1) \subseteq \ball(S_i,1)$ the feasibility of
the solution is preserved. Performing this replacement for all points in $R'
\cap U$ induces a feasible solution to \SETC/ of size no more than $|R'|$. It
follows that any $\alpha(n)$-approximate solver for \RDS/ yields an
$\alpha(n)$-approximation for \SETC/.
\end{proof}

We now prove Theorem~\ref{thm:hardassetcover}.

\begin{proof}[Proof of Theorem~\ref{thm:hardassetcover}]
As a corollary to Folklore~\ref{lem:setc-to-rds}, any polynomial-time
\triappp{(1-\eps)\ln(n)}{2-\eps'}{\cdot} for \RDS/ yields a polynomial-time
$((1-\eps)\ln(n))$-approximation for \SETC/ problem, by taking $P$ to be a
single level temporal-sampling with $P(1)=C$ and invoking it with $r=1$, $\delta
=0$, and successive values of $k$ until it succeeds in producing a clustering.
Since the first non-failing value of $k$ is at most the size of an optimum
solution to \SETC/, the resulting clustering is at most $(1-\eps)\ln(n)$ times
larger. The hardness now follows for any $\eps > 0$ by a result of Dinur and
Steurer \cite{Dinur14}.
\end{proof}

\subsection{Inapproximability in $2$-dimensional Euclidean space}
\label{sec:inapprox2}
While Theorem~\ref{thm:hardassetcover} shows that an increase in the number of
clusters should be expected if we demand to have a polynomial-time algorithm
that closely approximates the optimal radius and displacement, the construction
involves a somewhat unnatural metric space. We show that this condition remains
even for $2$-dimensional Euclidean space.

\begin{theorem}[Hardness of \MAX2SAT/\cite{Bellare98}]\label{thm:max2sgap}
There exist constants $0 < s < c < 1$, satisfying $c > 0.9$ and $c/s=74/73$,
such that it is \NP-hard to decide whether a given $2$-\CNF/ formula admits an
assignment which satisfies at least $c$-fraction of the clauses, or
whether any assignment satisfies at most $s$-fraction of clauses.
\end{theorem}

Let $\mathcal{S}$ be an instance of \MAX2SAT/ consisting of $l$ variables and
$m$ clauses. Given $0 < \delta_0 < r_0$, our goal is to construct
a temporal-sampling $P$ in polynomial-time such that:
\begin{enumerate}
\item $P$ admits a \triclusp{2m+(1-c)m}{r_0}{\delta_0} if there exists a
truth assignment that satisfies at least $c\cdot m$ clauses of $\mathcal{S}$.
\item $P$ does not admit a \triclusp{k}{\rho}{\delta_0} for any $k < 2m+(1-s)m$
and $\rho < 2 r_0$, if every truth assignment satisfies at most $s\cdot m$
clauses,
\end{enumerate}
where $c$ and $s$ are the constants from Theorem~\ref{thm:max2sgap}

\begin{figure}
\centering
\includegraphics[width=\linewidth]{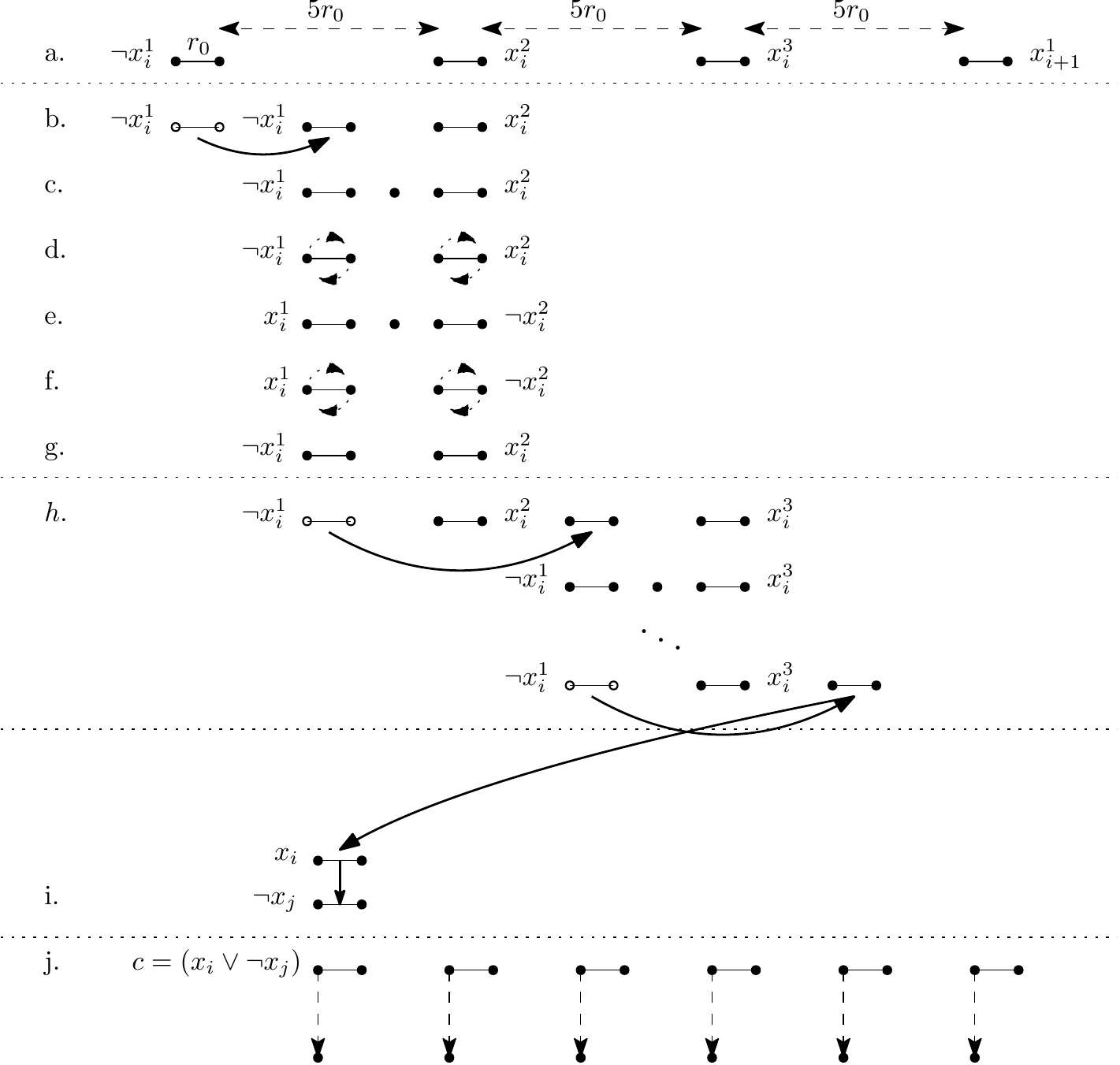}
\caption{\emph{Schematic overview of Phases 1 and 2}. Line $a$ shows an example
of the initial position of variable gadget copies corresponding to variable
$x_i$ and the first variable gadget copy of $x_{i+1}$. Lines $b$ through $g$ show
the consistency checking of \emph{Phase 1} as time progresses (downwards). In
line $b$, the variable gadget copy $x_i^1$ moves to align end-to-end with
$x_i^2$. Next an extra point appears mid-way between them for a single level
(line $c$). It disappears and then the gadgets rotate so that they are flipped
over (line $d$). An extra point then reappears (line $e$). In line $f$, both
gadgets flip back over, ending in the position on line $g$. In line $h$, this
process continues with $x_i^1$ moving to align with $x_i^3$, before again
performing the consistency check against $x_i^3$. This process goes on repeating
until $x_i^1$ has checked against $x_i^j$ for all $j > 1$. After this $x_1$ goes
on to assemble a clause $c=(x_i \vee \neg x_j)$ in line $i$ by overlapping with
a copy of a variable gadget for $x_j$ (already in place). The process then
repeats with $x_i^2$ checking against $x_i^j$ for $j > 2$. After all variable
gadget copies have been checked in this manner, all clauses have been assembled.
Each clause then receives an extra point, all of which simultaneously undergo a
motion perpendicular to each clause gadget, as shown in line $j$.
\label{fig:2satwhole}}
\end{figure}

\paragraph*{Variable gadgets.}
For each variable $x_i$ of $\mathcal{S}$, let $k_i$ be the number of literals
where it appears. We introduce $k_i$ pairs of points into each level of $P$,
which we denote by $x_i^j$ and $\neg x_i^j$ for $j \in [k_i]$. Our initial
arrangement of variable gadgets in the first level of $P$ is as in the top row
of Figure~\ref{fig:2satwhole}.

\paragraph*{Clause gadgets.}
We arrange a configuration of points for each clause $c=(l_1 \vee l_2)$ of
$\mathcal{S}$ by rigidly transporting one of the variable gadget copies of each
of its variables to a predetermined location. In doing so we overlay them so
that the points of each gadget corresponding to $l_1$ and $l_2$ overlap on one
side (see Figure~\ref{fig:2satwhole}$g$). The distance between two neighboring
clause gadgets is $4r_0$.

\begin{description}
\item{\textbf{Phase 1: Consistency checking and clause assembly.}} Each variable
in $\mathcal{S}$ corresponds to one or more variable gadgets (which we think of
as copies). We check each pair of variable gadget copies to ensure that all
copies have at least one side selected as a center. That is, either $x_i^j$ has
a center for all $j \in [k_i]$, or all $\neg x_i^j$ do (or both).  Initially, we
declare all variable gadget copies corresponding to the same variable as
``unchecked''. For each variable we perform the following procedure: The
unchecked copy with smallest index $j$ (the left most one, see
Figure~\ref{fig:2satwhole}) undergoes rigid motion so that it is aligned end to
end with a higher indexed copy $j' > j$. The two copies align with a distance of
$2r_0$ in-between. Moreover, the gadgets are consistently oriented with $\neg
x_i$ on the left and $x_i$ on the right. A consistency check is performed
between the pair consisting of the following steps: $1.$ An additional point is
introduced for one level at the mid-way point between the two gadgets. $2.$ It
subsequently disappears and the two variable gadgets simultaneously rotate in
place in the same direction about their respective midpoints so that they are
both oriented in with $x_i$ on the left. $3.$ The additional point reappears for
a single level. $4.$ After this the variable gadget copies rotate once more in
place so that $x_i$ is on the right. The $j$-th copy now undergoes rigid motion
to check against the next copy ($j'+1$-st, if it exists). After all $j$ has
checked with all $j' > j$, it is declared ``checked'' and  undergoes rigid
motion to take its place in the clause gadget. This process repeats until all
copies of the variable are checked. Finally, the lone remaining copy undergoes
rigid motion to its place in some clause gadget. The total traveled distance for
any copy of a variable gadget is $O(m r_0)$. Since it may only do so in steps of
size $\delta_0$, the total number of steps per copy is $O(m r_0/\delta_0)$ for a
total of $O(m^2 r_0/\delta_0)$ over all variable copies.
\item{\textbf{Phase 2: Satisfiability check.}} An extra point is introduced in
each clause gadget. For each clause $c=(l_1 \vee l_2)$, this occurs at the point
where the sides of the variable gadget copies corresponding to $l_1$ and $l_2$
have been identified. These points undergo motion along a line orthogonal to the
variable gadget copies for a distance of $4 r_0$ (see
Figure~\ref{fig:2satwhole}$h$). This phase takes $O(r_0 / \delta_0)$ steps.
\end{description}

\begin{lemma}\label{lem:highsatinst}
If there exists a truth assignment satisfying at least $c\cdot m$ clauses then
$P$ admits a \triclusp{2m+(1-c)m}{r_0}{\delta_0}.
\end{lemma}
\begin{proof}
Suppose such a truth assignment, $T$, exists. We use this assignment to
determine $2m$ of the $2m+(1-c)m$ cluster centers. For each variable $x_i$, if
$T(x_i)$ is \TRUE/ we select the points $x_i^j$ for all $j \in [k_i]$ as centers
in all levels up to and including the first \emph{Phase 2} level. Otherwise we
select all $\neg x_i^j$ for all $j \in [k_i]$ as centers in those same levels.
With these selections each satisfied clause has at least one center present at
the extra point of the clause gadget at the start of \emph{Phase 2}. Hence, for
satisfied clauses we arbitrarily select one of these centers and assign it to
the extra point for all remaining levels. For the unsatisfied clauses we assign
one of the remaining $(1-c)m$ centers. Note that the ends of the variable
gadgets that meet at the extra point of each unsatisfied clause correspond to
(thus far) unselected literals. We select one of those literals from each clause
to be a center in all levels up to and including the first \emph{Phase 2} level,
and select each unsatisfied clause's extra point in the remaining levels. The
remaining centers can be assigned arbitrarily to some $x_i^j$ in all levels. Let
the clustering induced by this choice of centers be denoted as $\mathcal{C}$.

We now argue that $\rad(\mathcal{C}) \leq r_0$. First note that for each
variable, $x_i$, for any $j \in [k_i]$ either the point $x_i^j$ or $\neg x_i^j$
is a center by assignment from $T$. Since the variable gadgets only undergo
rigid motion, any level of $P$ with only the points from the variable gadgets is
covered at a cost of at most $r_0$. We need only consider the levels of $P$
where additional points exist. Since the assignment from $T$ is consistent, it
follows that the extra points introduced during the consistency check involving
$x_i^j$ and $x_i^{j+1}$ for $j \in [k_i - 1]$ are always within a distance of
$r_0$ of their corresponding centers. For the levels in $\emph{Phase 2}$ all
extra points are covered and at least one center remains on each clause gadget,
which has diameter $r_0$.
\end{proof}

\begin{lemma}\label{lem:lowsatinst}
If every truth assignment satisfies at most $s\cdot m$ clauses then $P$ does not
admit a \triclusp{k}{\rho}{\delta_0} for $k < 2m+(1-s)m$, $\rho<2 r_0$
\end{lemma}
\begin{proof}
We argue that if $P$ admits \triclusp{k}{\rho}{\delta_0} for some $k <
2m+(1-s)m$, $\rho < 2 r_0$, then there exists a truth assignment that satisfies
more than $s\cdot m$ clauses. For any variable, all copies of its gadget in the
first level contain at least one center, as otherwise one of them is completely
uncovered and the nearest center is at least a distance of $2r_0$ away. Note
that the only points of a variable gadget which pass within a distance of
another $\delta_0$ are points which are eventually identified within a clause
gadget (just before \emph{Phase 2}). Thus, the choice of centers in the first
level completely determines the choice of centers at the start of \emph{Phase
2}. Further, this assignment is consistent in the sense that for each variable
$x_i$ either $x_i^j$ for all $j \in [k_i]$ are centers, or all of $\neg x_i^j$
for all $j \in [k_i]$ are (this is not mutually exclusive, some variable gadgets
might have both). To see why suppose that for some variable $x_i$ there is a
pair of copies corresponding to $j, j' \in [k_i-1]$ which disagree. In this case
each copy has exactly one center. Without loss of generality suppose $x_i^j$ and
$\neg x_i^{j'}$ have centers. There is a level where the extra point of the
consistency check is exactly midway between the opposite ends of these copies.
That is, the extra point is midway between $\neg x_i^j$ and $x_i^{j'}$. It
follows that the nearest centers to the extra point are at a distance of $2 r_0$
contradicting that $\rho< 2r_0$. Finally, note that the extra point of each
clause gadget is covered. This is only possible if at least one of the literals
that meet at the extra point have a center. Now consider the truth assignment
$T$, where $T(x_i)$ is \TRUE/ if and only if all $x_i^j$ for all $j \in [k_i]$
are centers in \emph{Phase 1}. Let $|T|$ denote the number of clauses satisfied
by $T$. We know that this assignment together with $u = k-2m$ additional centers
is sufficient to satisfy all clauses. That is, $|T| + u >= m$. Since there are
only $k < 2m + (1-s)m$ centers in total, it follows that the number of
additional centers required is strictly less than $(1-s)m$. Thus, $(1-s)m +|T| >
|T| + u >= m$, and we conclude that $|T| > s\cdot m$.
\end{proof}

\begin{lemma}\label{lem:convtime2sat}
Given an instance ${\cal S}$ of \MAX2SAT/ with $l$ variables and $m$ clauses, it
is possible to construct the above temporal-sampling of size $n \in O(m^3
r_0/\delta_0)$ in $O(m^3 r_0/\delta_0)$-time.
%In general we think of $r_0/\delta_0$ as a function of the input. Moreover, this construction is polynomial in the size of ${\cal S}$ whenever $r_0/\delta_0$ is.
%
\end{lemma}
\begin{proof}
This is immediate by construction.
\end{proof}

We are now ready to prove Theorem~\ref{thm:ccpolyhard}.

\begin{proof}[Proof of Theorem~\ref{thm:ccpolyhard}]
Let $\eps, \eps' \in \mathbb{R}$ with $\eps > 0$, $\eps' > 0$ be given. Let
$\omega > 6(1/\eps - 1)$. Let $\mathcal{S}$ be an instance of \MAX2SAT/ with $l$
variables and $m$ clauses which is promised to either admit a solution which
satisfies at least $c \cdot m$ clauses (high satisfiability) or does not admit
any solution which satisfies more than $s \cdot m$ clauses (low satisfiability),
for $c$ and $s$ from Theorem~\ref{thm:max2sgap}. Let $P$ be the result of the
above construction with $r_0/\delta_0 = l^\omega$. It follows from
Lemma~\ref{lem:convtime2sat} that the size of $P$ is $n \in O(l^{6+\omega})$
since $m \in O(l^2)$. From this and the choice of $\omega$ there exists a
universal constant $c_1$ such that $c_1 n^{1-\eps} \leq r_0/\delta_0 =
l^{\omega}$. Suppose there exists some \triappp{c_0}{2-\eps'}{c_\delta
n^{1-\eps}} for some $c_0, c_\delta \in \mathbb{R}$ with $c_\delta < c_1$. Run
the approximation on $P$ with $k=2m+(1-c)m$, $r=r_0$, $\delta = \delta_0$. Then
either it fails and Lemma~\ref{lem:highsatinst} implies there does not exist a
truth assignment satisfying at least $c \cdot m$ clauses (thereby deciding that
it is a low satisfiability instance in polynomial-time), or it outputs a
clustering $\mathcal{C}$. Since $c_\delta n^{1-\eps} < r_0/\delta_0$, it is easy
to see that Lemma~\ref{lem:highsatinst} and Lemma~\ref{lem:lowsatinst} still
apply. If $\mathcal{S}$ is a high satisfiability instance then it follows by
Lemma~\ref{lem:highsatinst} that the resulting clustering consists of at most
$c_0 \cdot (2m+(1-c)m)$ clusters with $\rad(\mathcal{C}) \leq (2-\eps')r_0$.
Otherwise by Lemma~\ref{lem:lowsatinst} it cannot consist of fewer than
$2m+(1-s)m$ centers with $\rad(\mathcal{C}) < 2r_0$. Thus it is possible to
distinguish high satisfiability instances from low satisfiability instances
provided that $(2-\eps') < 2$ and $c_0 \cdot (2m+(1-c)m) < 2m + (1-s) m$. The
former inequality holds by choice of $\eps'$. From the later inequality it
follows that the input classes will be distinguishable provided that $c_0 <
(3-s)/(3-c)$. The theorem follows by setting $c_0$ to be the infimum of the
right hand side subject to the constraints placed on $c$ and $s$ in
Theorem~\ref{thm:max2sgap}.
\end{proof}

\subsection{Inapproximability with exact number of clusters for \TKMED/}
\label{sec:1polypolyhardkm}
We remark here that the construction of section~\ref{sec:nphardsmalldelta} which
demonstrates \NP-hardness of $(1, \poly(n), \poly(n))$-approximation works for
\TKMED/ with only minor modification.

Given any positive real numbers $\eps$, $s \leq 1$, $r_0$, and $\delta_0 < r_0
\sqrt{3}/4$, and some $l$-variable instance ${\cal S}$ of {\textsc Exact-3-SAT},
using essentially the same construction as used in
section~\ref{sec:nphardsmalldelta} we show how to construct a temporal sampling
$P$ such that the following conditions hold:
\begin{enumerate}
  \item $P$ admits a \triclusp{l}{\frac{1}{2}r_0(l+1)}{\delta_0} if ${\cal S}$ is satisfiable.
  \item $P$ does not admit a \triclusp{l}{c_rn^{s(1-\eps)}r_0}{c_\delta n^{(1-s)(1-\eps)}\delta_0} otherwise.
\end{enumerate}
The only difference will be in our selection of the value of the constant $\rho$,
which we determine later.

\begin{lemma}\label{lem:satkm}
If ${\cal S}$ is satisfiable then $P$ admits an
\triclusp{l}{\frac{1}{2}r_0(l+1)}{\delta_0}.
\end{lemma}
\begin{proof}
We make the identical assignments as Lemma~\ref{lem:sat}, the only difference
here is that we use the $k$-median objective. The cost of covering the variable
gadgets in \emph{Initial Position}, \emph{Phase 1}, and \emph{Phase 3} are
$\frac{1}{2}r_0l$. In \emph{Phase 2} one of the remaining centers near where the
clauses meet needs to cover the points of the variable gadget which gives its
center to the extra point. The cost of this level is $\frac{1}{2}r_0(l+1)$.
\end{proof}

\begin{lemma}\label{lem:unsatkm}
If ${\cal S}$ is not satisfiable then $P$ does not admit an
\triclusp{l}{\frac{1}{2}r_0l + \rho r_0}{\delta_0}.
\end{lemma}
\begin{proof}
The proof is identical to Lemma~\ref{lem:unsat}, except we use the $k$-median
objective and require that $\rho > l+1$ so that the spacing of the variable
gadgets in initial position exceeds the cost of their $k$-median clustering.
\end{proof}

Since the construction is the same as in section~\ref{sec:nphardsmalldelta}
up to the choice of $\rho$, we can apply Lemma~\ref{lem:convtime} to conclude
$|P| \in O(\rho r_0l^2 m / \delta_0)$.

We are now ready to prove Theorem~\ref{thm:1polypolyhardkm}.

\begin{proof}[Proof of Theorem~\ref{thm:1polypolyhardkm}]
Let $c > 5(1/\eps - 1)$. Let ${\cal S}$ be an instance of \THREESAT/ with $l$
variables and $m$ clauses. We invoke the construction of
section~\ref{sec:nphardsmalldelta} with $\rho(l)=l^{s\cdot c}$,
$r_0(l)/\delta_0(l) = l^{(1-s)\cdot c}$ and yield, in polynomial-time, a
temporal-sampling $P$ over $\mathbb{R}^2$ of size $n \in O(l^{c+5})$. Note here
we have used the fact that $m \in O(l^3)$. Suppose the existence of some
polynomial-time \triappp{1}{\alpha(n)}{\beta(n)} where $\alpha(n) = c_r
n^{s\cdot(1-\eps)}$, and $\beta(n) = c_\delta n^{(1-s)(1-\eps)}$. There exists a
universal constant $c_1$ (where $c_1 > 1$ by construction) such that $c_1
n^{(1-s)\cdot(1-\eps)} \leq (\sqrt{3}/4) r_0/\delta_0 = (\sqrt{3}/4)
l^{(1-s)\cdot c}$. Thus we let $1 \leq c_\delta < c_1$, As with the remark of
section~\ref{sec:nphardsmalldelta}, Lemma~\ref{lem:satkm} and
Lemma~\ref{lem:unsatkm} still apply. We run this approximation on $P$ with
$k=l$, $r=\frac{1}{2}r_0(l+1)$, and $\delta=\delta_0$. In polynomial-time the
algorithm either produces output or fails. If the algorithm fails it follows by
Definition~\ref{def:APPROXTRICLUS} that $P$ does not admit a
\triclusp{l}{\frac{1}{2}r_0(l+1)}{\delta_0} and thus ${\cal S}$ is not
satisfiable by Lemma~\ref{lem:satkm}. Otherwise, the algorithm produces some
clustering ${\cal C}$ of radius $\mediancost(\mathcal{C}) \leq \alpha(n)
\frac{1}{2}r_0(l+1)$. Lemma~\ref{lem:unsatkm} ensures that if $\mathcal{S}$ is
not satisfiable then $\mediancost(\mathcal{C}) \geq\frac{1}{2}r_0l + \rho r_0$.
Thus provided that $\alpha(n) \frac{1}{2}r_0(l+1) \leq \frac{1}{2}r_0l + \rho
r_0$, the algorithm produces output if and only if ${\cal S}$ is satisfiable,
giving a polynomial-time test for satisfiability. That is, whenever \[ \alpha(n)
\leq \frac{l + 2l^{s\cdot c}}{l+1} < 3l^{s \cdot c} \leq c_r n^{\frac{s \cdot
c}{c+5}} \leq c_r n^{s \cdot (1-\eps)}, \] for some constant $c_r$.
\end{proof}

\subsection{\NP-hardness of $(O(1), O(1), \poly(n))$-approximation for \TKMED/}
\label{sec:ccpolyhardkm}
In this section we show that it is is \NP-hard to simultaneously approximate
both the number of clusters and the spatial cost arbitrarily well.

\begin{definition}\label{def:expander}
A $d$-regular graph $G=(V,E)$ is an $\omega$-expander if for every set $S
\subset V$ where $|S| \leq \frac{1}{2} |V|$ at least $\omega d |S|$ edges
connect $S$ and $V \setminus S$.
\end{definition}

Let $\mathcal{S}$ be an instance of \MAX2SAT/ consisting of $l$ variables and
$m$ clauses. Given $0 < \delta_0 < r_0$, our goal is to construct a
temporal-sampling $P$ in polynomial-time such that:
\begin{enumerate}
\item $P$ admits a \triclusp{2m+(1-c)m}{7mr_0}{\delta_0} if there exists a
truth assignment that satisfies at least $c\cdot m$ clauses of $\mathcal{S}$.
\item $P$ does not admit a \triclusp{k}{\rho}{\delta_0} for any $k <
2m+(1-s)m-fm$ and $\rho < (7 + c_2 f)mr_0$, for some universal constant $c_2
\geq 0$, and any $0 \leq f \leq 1/2$, if every truth assignment satisfies at
most $s\cdot m$ clauses.
\end{enumerate}
Here, $c$ and $s$ are the constants from Theorem~\ref{thm:max2sgap}.

\paragraph*{Variable gadgets.}
For each variable $x_i$ of $\mathcal{S}$, let $k_i$ be the number of literals
where it appears. We introduce $k_i$ pairs of points into each level of $P$,
which we denote by $x_i^j$ and $\neg x_i^j$ for $j \in [k_i]$. We set the
distance $d(x_i^j,\neg x_i^j)=r_0$ for all $j \in [k_i]$ in all levels. We think
of each pair as a variable gadget for the variable $x_i$ so that any level
contains $k_i$ ``copies'' of this gadget.

\paragraph*{Clause gadgets.}
We arrange a configuration of points for each clause $c=(l_1 \vee l_2)$ of
$\mathcal{S}$ by transporting one of the variable gadget copies of each of its
variables to a predetermined location. Ultimately, we overlay them so that the
points of each gadget corresponding to $l_1$ and $l_2$ overlap on one side,
with $\neg l_1$, $\neg l_2$ overlapping on the other.

\begin{figure}
\begin{center}
\includegraphics[width=0.36\textwidth]{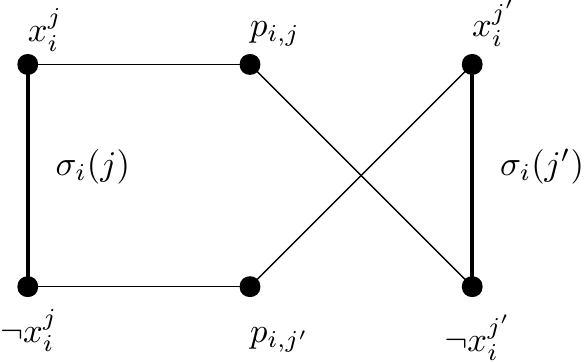}
\end{center}
\caption{\label{fig:max2sat-km-consistency}
All lines indicate a distance of $r_0$. Dark lines join the points of variable
gadgets $\sigma_i(j)$, $\sigma_i(j')$, respectively.
}
\end{figure}

\begin{description}
\item{\textbf{Phase 1: Initial layout of $P(1)$.}} Fix an $\omega$. To determine
the initial distances among the variable copies we generate a $k_i$-vertex
$3$-regular $\omega$-expander, $G_i=(V_i, E_i)$ for each variable $x_i$. Fix
bijections $\sigma_i$ from $V_i$ to the set of copies of variable gadgets for
$x_i$ in $P(1)$. For each edge $e=(j,j') \in E_i$ we introduce a pair of
auxiliary points $p_{i,j}$, $p_{i,j'}$ into $P(1)$ and set the distances between
these points and the points of variable gadgets $\sigma_i(j)$, and
$\sigma_i(j')$ according to Figure~\ref{fig:max2sat-km-consistency}. For points
of variable gadget copies from distinct variables $x_i$, $x_{i'}$, we set
$d(x^j_i, \neg x^{j'}_{i'}) = d(\neg x^j_i, x^{j'}_{i'}) = (7+c_2)mr_0$ for
some to-be-determined constant, $c_2$. We think of this process of specifying
distances as building a weighted graph on the points of $P(1)$, where the
weights are given by the distance values. Thus we allow any remaining
unspecified distances of $P(1)$ to be given by the shortest paths distance in
this graph.
\item{\textbf{Phase 2: Isolation of variable gadgets.}} In $P(1)$ all variable
gadget copies are separated by distances of at least $2r_0$. In this phase we
simultaneously expand the distances between the points of any pair of variable
gadget copies to $(7+c_2)mr_0$ in steps of size at most $\delta_0$. The only
points present in these levels are the $2m$ points of the variable gadget
copies. The number of levels required for this stage is
$O(mr_0/\delta_0)$
\item{\textbf{Phase 3: Condensation of clause gadgets.}} For each clause $(l_i
\vee l_{i'})$. Let $x_i$, $x_{i'}$ denote the variables referenced by literals
$l_i$, and $l_i'$, respectively. We select $j \in [k_i]$, $j' \in [k_{i'}]$
corresponding to yet unused copies of variable gadgets $x_i$ and $x_i'$,
respectively. By a slight abuse of notation we label the points corresponding to
$l_i$ and $l_{i'}$ with the labels $l_i$, $l_{i'}$. In other words $l_i$
corresponds to $\neg x^j_i$ if it is negated, otherwise $x^j_i$ with $l_{i'}$
defined analogously. We transport each variable gadget so that the points of
each which correspond to $l_i$ and $l_{i'}$ overlap on one side, with $\neg
l_i$, $\neg l_{i'}$ overlapping on the other. We call the point where $l_i$ and
$l_{i'}$ overlap the \emph{location} of the clause. The number of levels
required for this stage is $O(mr_0/\delta_0)$
\item{\textbf{Phase 4: Clause verification.}} Once the clause gadgets have been
assembled, we introduce an extra point at the location of each of clause. These
extra points move directly away from their respective clauses for a distance
of $(7+c_2)mr_0$. The number of levels required for this stage is
$O(mr_0/\delta_0)$
\end{description}

Let $P$ denote the temporal-sampling given by the above
construction.

\begin{lemma}\label{lem:kmsat}
If there exists a truth assignment satisfying at least $c\cdot m$ clauses,
then $P$ admits a \triclusp{2m+(1-c)m}{7mr_0}{\delta_0}.
\end{lemma}
\begin{proof}
Let $T$ be a satisfying assignment which satisfies at least $c$-fraction of the
clauses. We now use $T$ to construct a clustering $\mathcal{C}$ with
$|\mathcal{C}| \leq 2m+(1-c)m$, $\delta(\mathcal{C}) \leq
\delta_0$. Introduce $2m$ trajectories $\tau_{i,j}$ such that
\[
\tau_{i,j}(i') =
  \begin{cases}
             x_i^j & \text{if } T(x_i) = \TRUE/\\
        \neg x_i^j & \text{if } T(x_i) = \FALSE/\\
  \end{cases},
\]
for all $i \in [l]$, for all $j \in [k_i]$, and for all $i'$ starting at the
first level of \emph{Phase}~$1$ through the last level of \emph{Phase}~$3$. Note
that since $T$ satisfies $\mathcal{S}$, every satisfiable clause has at least
one trajectory at its \emph{location} at the end of \emph{Phase}~$3$. We extend
one such trajectory to cover the clause's extra point throughout \emph{Phase}~4.
The other trajectory which is near to the clause maintains the value that it had
at the end of \emph{Phase}~3 through all levels of \emph{Phase}~4. This assigns
$2m$ of the at most $2m+(1-c)m$ trajectories. We will now use the remaining
$(1-c)m$ to pay for the unsatisfiable clauses. For each clause gadget
corresponding to some unsatisfied clause, select one of the literals $l_i^j$ at
its location (that is, $l_i^j$ is either $x_i^j$ or $\neg x_i^j$ for some $i \in
[l]$, $j \in [k_i]$). We introduce a trajectory $\tau$ into the clustering such
that $\tau = l_i^j$ from the first level of \emph{Phase}~$1$ through the last
level of \emph{Phase}~$3$, and which covers the clause's extra point throughout
each level of \emph{Phase}~$4$. This completes our assignment. Since there are
no more than $(1-c)m$ unsatisfied clauses, the size of the clustering is at most
$2m+(1-c)m$, as desired. Further, by construction the displacement between
trajectory centers on adjacent levels is at most $\delta_0$. Thus
$\delta(\mathcal{C}) \leq \delta_0$.

We now argue that $\mediancost(\mathcal{C}) \leq 7mr_0$. Note that in
all levels, all points are within a distance of $r_0$ from some center. In
particular, by assignment from $T$, it holds that for any $i$ either $x_i^j$ has
a center for all $j \in [k_i]$ or $\neg x_i^j$ has a center for all $j \in
[k_i]$ for all levels of the first three phases. Thus each variable gadget has a
center at at least one end, so we can cover all variable gadget points at a cost
of $mr_0$. This same cost of covering the variable gadgets also holds in
\emph{Phase}~$4$, as at least one trajectory remains incident to the overlapping
pair of variable gadgets. Since the second through last levels of
\emph{Phase}~$1$, all levels of \emph{Phase}~2, and all levels of \emph{Phase}~3
only contain points which correspond to variable gadgets, these levels be
covered at a cost of $mr_0$. This bound also extends to the points of
\emph{Phase}~4 since any extra points which appear in those levels are centers
of some trajectory by construction. It only remains to bound the cost of $P(1)$.
Recall that there are pairs of extra points in $P(1)$ which correspond to edges
of each expander. The fact that all of either the \TRUE/ or \FALSE/ sides of
each variable gadget receive a center, implies that all extra points are at a
distance of $r_0$ to a center. Thus the cost of covering the first level is at
most $mr_0+2r_0\sum_{i \in [l]}|E_i|$. Since each $G_i$ is $3$-regular, $|E_i|
\leq \frac{3}{2} |V_i| = \frac{3}{2} k_i$. Thus, $mr_0 + 2r_0\sum_{i \in
[l]}|E_i| \leq mr_0 + 3 r_0 \sum_{i \in [l]} k_i = 7mr_0$.
\end{proof}
\begin{lemma}\label{lem:kmunsat}
If every truth assignment satisfies at most $s\cdot m$
clauses then, for any $0 \leq f < 1$, $P$ does not admit a
\triclusp{2m+(1-s)m-fm}{(7+c_2 f)mr_0}{\delta_0}.
\end{lemma}
\begin{proof}
Let $0 \leq f < 1$. We will argue that if $P$ admits a
\triclusp{k}{\rho}{\delta_0}, $\mathcal{C}$, for some $k < 2m+(1-s)m-fm$, and
$\rho < (7+c_2 f)mr_0$, then there exists a truth assignment which satisfies
more than $s\cdot m$ clauses. First note that the distance from any extra point
of $P(1)$ to its nearest neighbor in $P(2)$ is $r_0$. Thus, since $\delta_0 <
r_0$, no such point has a feasible successor. It follows that none of these
points appear in the trajectories of $\mathcal{C}$. Further, there is a unique
valid successor for all variable gadget points in all levels of
\emph{Phase}~$1$, \emph{Phase}~$2$, and all but the later the levels of
\emph{Phase}~$3$ when the corresponding ends of the overlaid variable gadgets
come within $\delta_0$ of each other. Since $\mediancost(\mathcal{C})
\leq \rho < (7+c_2f)mr_0$, it must be the case that the farthest distance to
any cluster center is within $(7+c_2f)mr_0$. In particular this means that every
variable gadget has been assigned a center. As otherwise some variable gadget at
the end of \emph{Phase}~$2$ is covered by a center of another and the nearest
one is at a distance of $(7+c_2)mr_0$ away. Further, since the extra point of
\emph{Phase}~$4$ also moves a distance of $(7+c_2)mr_0$, it must also be the
case that some trajectory is incident to the location of each clause.

We would now like to infer a truth assignment from $\mathcal{C}$, but the main
obstacle is that both $x_i^j$ and $\neg x_i^{j'}$ can be trajectories. We will
show that the total discrepancy is bounded, and that even after accounting for
``unfairly'' satisfied clauses, it is possible to construct a satisfying
assignment which satisfies more than $s\cdot m$ clauses. To see this, note that
for each pair of trajectories corresponding to the same variable which disagree
and share an edge in $G_i$, the gadget in
Figure~\ref{fig:max2sat-km-consistency} costs an additional $r_0$ to cover. For
each $G_i$ let $I_i$ denote the subset of vertices of $V_i$ which correspond to
the minority assignment. That is, $I_i$ is the subset of vertices of $V_i$ which
correspond to the subset of either \TRUE/ or \FALSE/ variable gadgets, whichever
has smaller cardinality. It follows that $\frac{1}{m}\sum_{i \in [l]}|I_i|$ is
an upper bound on the total fraction of clauses satisfied by inconsistent
trajectories. The additional coverage cost of the inconsistencies is equal to
the number of edges which cross the $(I_i, V_i \setminus I_i)$-cut, which, since
each $G_i$ is a $3$-regular $\omega$-expander, is at least $3\omega|I_i|$. Since
$\mediancost(\mathcal{C}) \leq (7+c_2f)mr_0$, we have that $7mr_0 +
\frac{3\omega}{m}\sum_{i \in [l]}|I_i| mr_0 \leq (7+c_2f)mr_0$. Thus,
$\frac{3\omega}{m}\sum_{i \in [l]}|I_i| \leq c_2f$. By setting $c_2=3\omega$,
$f$ is an upper bound on the fraction of clauses flipped by inconsistent
trajectories. Let $I:P(1)\rightarrow\{\TRUE/,\FALSE/\}$ such that
\[
I(x_i^j)=
  \begin{cases}
             \TRUE/  & \text{if } x_i^j \text{ has a center.}\\
             \FALSE/ & \text{if } \neg x_i^j \text{ has a center.}\\
  \end{cases}
\]
Consider a truth assignment $T$ such that $T(x_i)=\maj_{j \in [k_i]}(I(x_i^j))$,
and let $|T|$ denote the number of clauses satisfied by $T$. Since
$\mediancost(\mathcal{C}) \leq (7+c_2f)mr_0$ it must be the case the extra points
for all clauses are covered. That is, $T$ together with $u = k-2m$ additional
trajectories is sufficient for satisfying all clauses, so that $|T| + u \geq m$.
Since $k < 2m + (1-s)m - fm$ we have that $u < (1-s)m - fm$. Combining both
inequalities we see that $(1-s)m - fm + |T| > |T| + u \geq m$. It immediately
follows that $|T| > (s+f)m \geq s\cdot m$. \end{proof}

We are now ready to prove Theorem~\ref{thm:ccpolyhardkm}.

\begin{proof}[Proof of Theorem~\ref{thm:ccpolyhardkm}]
Let $\eps > 0$ be given and suppose that there exists a polynomial-time
$(\alpha, \beta, c_\delta n^{1-\eps})$-approximation. Let $\mathcal{S}$ be an
instance of \MAX2SAT/ which is promised to either admit an assignment which
satisfies at least $c$ fraction of the clauses, or does not admit an assignment
which satisfies more than $s$ fraction. Let $0 \leq f < c-s$. Construct $P$ from
$\mathcal{S}$ in $O(m^2 r_0/\delta_0)$ time. Note that we are free to take
$\delta_0$ in the construction of $P$ as small as we like, provided that the
number of levels of $P$ remains polynomial in the size of $\mathcal{S}$. Using
this freedom we take $\delta_0$ such that $c_\delta n^{1-\eps} < r_0/\delta_0$.
Run this approximation on $P$ with $k=2m+(1-c)m$, $r=7mr_0$, $\delta=\delta_0$.
If the approximation fails then it must be the case that $\mathcal{S}$ is a \NO/
instance. Otherwise, the approximation produces a clustering, $\mathcal{C}$,
with at most $\alpha(2m+(1-c)m)$ trajectories, $\mediancost(\mathcal{C})\leq \beta
7mr_0$, and $\delta(\mathcal{C}) \leq c_\delta n^{1-\eps}
\delta_0$. Note that Lemma~\ref{lem:kmunsat} and Lemma~\ref{lem:kmsat} still
hold since $c_\delta n^{1-\eps} \delta_0 < r_0$. Thus, provided that $\alpha
(2m+(1-c)m) < 2m+(1-s)m-fm$, and $\beta(7mr_0) < (7+3\omega f)mr_0$ it is
possible to distinguish \YES/ \MAX2SAT/ instances from \NO/ instances. That is,
for $\alpha < \frac{3-(s+f)}{3-c}$ and $\beta < 1 + \frac{3}{7}\omega f$. Taking
$c_r = \frac{3}{7}\omega$ completes the proof.
\end{proof}

\section{Conclusion}
Our results show that in many cases temporal clustering problems are hard to
approximate. On the other hand, our polynomial time algorithms show that in some
cases if we allow approximations in terms of parameters like $r / \delta$ or the
spread $\Delta$, the approximation becomes tractable. We wish to better
understand the boundary between these cases. Another
direction comes from altering the model. We currently
consider trajectories consisting of points in the input; an alternative
formulation could allow centers from the ambient metric space.
We plan to investigate this model in future research.

\section*{Acknowledgements}
This work was partially supported by the NSF grants CCF 1318595, CCF 1423230,
DMS 1547357, and NSF award CAREER 1453472.

\bibliographystyle{plain}
\bibliography{bibfile}

\end{document}